\documentclass[12pt,a4paper]{iopart}
\usepackage{bm}
\usepackage{epsfig}

\usepackage{bm}
\usepackage{epsfig}
\usepackage{amssymb}

\newcommand{\dd}{\mathrm{d}}
\newcommand{\ee}{\mathrm{e}}
\newcommand{\tobs}{t_\mathrm{obs}}

\newcommand{\sig}{\sigma}

\newcommand{\CC}{\mathcal{C}}
\newcommand{\WW}{\mathbb{W}}
\newcommand{\HH}{\mathbb{H}}

\newcommand{\from}{\leftarrow}

\newcommand{\nbar}{\overline{n}}

\begin{document}
\title[Large deviations of the dynamical activity in the East model]  
{Large deviations of the dynamical activity in the East model: analysing structure in biased trajectories} 
\author{Robert L. Jack}
\address{Department of Physics, University of Bath, Bath BA2 7AY, United Kingdom}
\author{Peter Sollich}
\address{Department of Mathematics, King's College London, Strand, London WC2R 2LS, United Kingdom}

\begin{abstract}
We consider large deviations of the dynamical activity in the East model.  We bias this system to larger than average
activity and investigate the structure that emerges.  To best characterise this structure, we exploit the
fact that there are effective interactions that would reproduce the
same behaviour in an equilibrium system.
We combine numerical results with linear response theory and
variational estimates of these effective interactions, giving the
first insights into such interactions in a many-body system, across a wide range of biases.   
The system exhibits a hierarchy
of responses to the bias, remaining quasi-equilibrated on short length scales, but deviating far from equilibrium on large length scales.  
%
%
We discuss the connection between this hierarchy and the hierarchical aging behaviour of the system.
\end{abstract}
\maketitle


\section{Introduction}

What is the probability that an extensive observable in a physical system has a value far from its average?  
Such questions are the subject of large-deviation theory~\cite{Touchette}, which provides a mathematical foundation for
classical thermodynamics~\cite{Ruelle}, and has more recently been applied to a variety of problems in non-equilibrium
statistical mechanics~\cite{fluct,JL,Bod-Der,Simon09,Imparato09,popkov,evans04,kcm-transition,hedges09}.
Notable results in these non-equilibrium settings include analyses of fluctuation
theorems~\cite{fluct}; exact results for exclusion processes and models of energy transport~\cite{JL,Bod-Der,Simon09,Imparato09,popkov}; 
a proposed non-equilibrium counterpart of detailed balance in sheared systems~\cite{evans04}; and 
dynamical phase transitions in glassy systems~\cite{kcm-transition,hedges09,jack-rom09,elmatad10}.  
Here, we focus on trajectories in which an extensive
measurement of dynamical activity~\cite{kcm-transition,maes06-activity,lecomte-chaotic} has a non-typical value, and we discuss how these trajectories can be characterised
in terms of effective interactions~\cite{Simon09,popkov,evans04,baule-prl08,simha-evans08,Jack-Sollich-PTP,baule-jstat2010,touchette-arxiv2013}.

The nature of these effective interactions is important in interpreting measurements of large deviations
in non-equilibrium systems.
In particular, if the interactions are simple
and short-ranged, one can interpret the large-deviation behaviour of the system in terms of its response to 
these short-ranged forces.  In this situation, physical intuition and results from the existing literature can be
very useful.  However, there are at least
some cases~\cite{JL,Bod-Der,elmatad10,Jack-Sollich-PTP} where long-ranged effective interactions appear, and 
lead to unusual new behaviour (for example, phase transitions to states 
with long-ranged order may appear in one-dimensional systems).  In these cases, interpreting and analysing
results for large deviations is more difficult.  An example is given by the very stable ``glass'' states
that appear in glassy model systems, when one considers large deviations of the dynamical 
activity~\cite{kcm-transition,hedges09,jack-rom09,jack11-stable,Speck12} --
it is not clear what effective interactions might be required to stabilise these states, and it is therefore
difficult to understand what kinds of equilibrium or non-equilibrium protocols might be used to prepare them 
in the laboratory.

In this study, we present numerical and analytical results for the one-dimensional East model~\cite{Jackle91}.  
This simple spin system is an example of a kinetically constrained model~\cite{Ritort-Sollich,GST-kcm},
where complex dynamical behaviour arises at low temperature, while all thermodynamic quantities remain very simple.
The model has been studied extensively in the context of glassy
systems~\cite{Sollich-Evans,Aldous02,East-gap}, particularly within the theory
of dynamical facilitation~\cite{GC-east}.   At low temperatures, it supports 
a very broad spectrum of time scales -- this results in a range of glassy phenomena, and
also considerable structure in the large deviation functions of the model.
The model is useful for our purposes because its large deviations are quite rich and complex,
but it is still tractable both numerically and analytically.  We present several methods for
characterising the effective interactions associated with large deviations in this model.
The one-dimensional nature of the model greatly facilitates our analysis in this article, but we argue that our methods
and general results  have potential applicability for analysing effective interactions in a wide
range of systems. Our results are the first to give direct insights
into effective interactions over a range of biases in a many-body system, going beyond
previously studied cases where very few degrees of
freedom were considered~\cite{popkov,evans04,baule-prl08,simha-evans08,baule-jstat2010,touchette-arxiv2013},
or the analysis was restricted to the limit of strong biasing~\cite{popkov}.  

The paper is organised as follows: In Section~\ref{sec:model}, we define the East model,
the biased ensembles of trajectories that we study, and the observables that we use to characterise these ensembles.
Section~\ref{sec:num} gives an overview of the response to the bias, illustrated by numerical results.
In Section~\ref{sec:pert}, we use a linear-response (perturbative) formalism 
to analyse the effective interactions in the system, for small $\nu$: the resulting physical picture
is discussed in Section~\ref{sec:hier}.  Since strong long-ranged effective interactions 
appear even at the perturbative level, we then use non-perturbative variational schemes to estimate
effective interactions (Sec.~\ref{sec:var}).  Finally, we discuss implications of our results
for more general systems in Sec.~\ref{sec:summ}.

\section{Model and ensemble definitions}
\label{sec:model}

\subsection{East model}

The one-dimensional East model~\cite{Jackle91,Ritort-Sollich} has binary spins $n_i=0,1$ where the sites $i=1\dots N$ form a linear chain, with periodic
boundaries.  We identify $n_i=1$ as the `up' state and $n_i=0$ as the `down' state.
The energy of the system has the simple form $E_0=\sum_i n_i$,
 and spins flip with Glauber rates, subject to the kinetic constraint that spin $i$ may
flip only if spin $i-1$ is in the `up' state, $n_{i-1}=1$.  That is,
spin $i$ flips from $0\to1$ with rate $n_{i-1}c$, and from $1\to0$ with rate $n_{i-1}(1-c)$, where $c=(1+\ee^\beta)^{-1}$ is equal
to the equilibrium fraction of `up' spins, and $\beta$ is the inverse temperature.
We use the notation
$\CC=(n_1,n_2,\dots,n_N)$ to represent a configuration of the system.  

We focus on the behaviour for small $c$ (low temperature).
At equilibrium, the dynamical behaviour of the system is hierarchical~\cite{Sollich-Evans}: 
motion on a length scale $\ell$ requires the system
to overcome an energy barrier of height
\begin{equation}
  \alpha_\ell = \lceil \log_2 \ell \rceil.
\label{equ:alpha}
\end{equation}
That is, $\alpha_\ell$ is the smallest integer greater than or equal to $\log_2 \ell$.
Specifically, this is the energy barrier associated with relaxation of an up-spin that 
is a distance $\ell$ from the nearest up-spin to its left.
At low temperature, the rates for such processes scale as 
\begin{equation}
  \tau_\ell = c^{\alpha_\ell}\sim \ee^{-\beta \alpha_\ell}, \qquad \ell \lesssim 1/c .
\label{equ:tau-ell}
\end{equation}
The typical distance between up spins in the system at equilibrium is $1/c$: 
if one assumes that (\ref{equ:tau-ell}) applies on this length scale then
 one estimates the relaxation time of the system to be $\tau_{1/c} \sim \ee^{\beta^2 /\ln 2}$~\cite{Sollich-Evans,Aldous02}.  However, 
 while this argument is persuasive, the bulk
relaxation time at equilibrium in fact scales as $\tau_0 \sim
\ee^{\beta^2 /(2\ln 2)}$~\cite{East-gap}. The slower divergence arises
because the simple argument above only considers the energy barrier of
the most efficient path, but neglects speed-ups arising from the
availability of many paths. These become significant for $\ell\approx 1/c$.
For length scales $\ell$ larger than $1/c$, the system relaxes by undergoing approximately
$c\ell$ successive events, each operating on a length scale of order $1/c$, and taking a time of order $\tau_0$: hence we expect $\tau_\ell \sim c\ell \tau_0$.

\subsection{Large deviations}
\label{sec:large-dev}

To investigate large deviations in the East model, we begin with its master equation:
\begin{equation}
\partial_t P(\CC,t) = -r(\CC) P(\CC,t) + \sum_{\CC'(\neq \CC)} W(\CC \from \CC') P(\CC',t) ,
\end{equation}
where $W(\CC\from\CC')$ is the transition rate from $\CC'$ to $\CC$ and
\begin{equation}
r(\CC) = \sum_{\CC' \neq \CC} W(\CC' \from \CC)
\label{equ:def-r}
\end{equation}
is the escape rate from configuration $\CC$.
Writing $|P(\CC,t)\rangle=\sum_\CC P(\CC,t)|\CC\rangle$, the master equation is
$\partial_t |P(\CC,t)\rangle = \WW |P(\CC,t)\rangle$ where $\WW$
is the `master operator' (or generator).  Using a spin-$\frac12$ basis for the configurations of the system, one has~\cite{kcm-transition}
\begin{equation}
\WW = \sum_i \hat{n}_{i-1} \left[ (1-c)\sig^-_i + c \sig^+_i - (1-2c)\hat{n}_i -c   \right] ,
\end{equation}
where $\hat{n}_i|\CC\rangle = n_i |\CC\rangle$ gives the state of spin $i$ in configuration $\CC$,
 while the $\sig_i^\pm$ are raising and lowering operators for spin $i$.  
%
%
(Explicitly, if we write a configuration with $n_i=0$ or $n_i=1$ as $|\cdots0\cdots\rangle$ or $|\cdots1\cdots\rangle$ respectively, then the operators act as $\sigma_i^+ |\cdots0\cdots\rangle= |\cdots1\cdots\rangle$ and $\sigma_i^- |\cdots1\cdots\rangle
 = |\cdots0\cdots\rangle$, while $\sigma_i^+|\cdots1\cdots\rangle = 0 = \sigma_i^-|\cdots0\cdots\rangle$.)

Large deviations of the dynamical activity in the East model were first
considered in~Ref.~\cite{Merolle} and later in~\cite{kcm-transition,elmatad13}. 
Consider a trajectory of length (``observation time'')
$\tobs$, which contains a total of $K$ configuration changes.  
%
%
At equilibrium, $K$ has a probability distribution
$P_0(K)$: if the system size $N$ and the observation time $\tobs$ are large then the central limit theorem implies
that the variance and mean of $K$ both scale as $N\tobs$.  Thus, $P_0(K)$ becomes sharply peaked as $N,\tobs \to\infty$.
Nevertheless, one may still consider trajectories with non-typical values of $K$.  One expects
\begin{equation}
P_0(K) \simeq \ee^{-N\tobs \pi_K(k)} ,
\end{equation}
where $k=K/(N\tobs)$ and $\pi_K(k)$ is a `spacetime free energy' or `rate function' that determines the probability of particular
values of $K$~\cite{pi-foot}.
%
%

In practice, it is convenient to follow an equivalent route, concentrating not on $\pi_K(k)$ but on its Legendre transform
$\psi_K(s) = \min_k[sk + \pi_K(k)]$.   To achieve this, one defines (as in Ref.~\cite{kcm-transition}) a probability distribution over trajectories: 
\begin{equation}
\mathrm{Prob}[\CC(t),s] = \mathrm{Prob}[\CC(t),0] \cdot \frac{ \ee^{-sK} }{ \langle \ee^{-sK} \rangle_0 }  ,
\label{equ:s-ens}
\end{equation}
where $\CC(t)$ represents a trajectory of the system and $K$ denotes its activity, while
the parameter $s$ biases the equilibrium distribution $\mathrm{Prob}[\CC(t),0]$, favouring trajectories with
non-typical values of $K$.  The notation $\langle \cdot \rangle_0$ indicates an equilibrium average. 

Following~\cite{kcm-transition,hedges09}, we refer to the probability distribution $\mathrm{Prob}[\CC(t),s]$ as an `$s$-ensemble'.
In the following,
we use $\langle \cdot \rangle_s$ to represent an average with respect to this distribution.
Standard arguments based on equivalence of ensembles indicate that averages with respect to $\mathrm{Prob}[\CC(t),s]$
are equivalent to averages over trajectories with fixed values of $K$ (see also~\cite{touchette-arxiv2013}).
In the long time limit, the free energy $\psi_K(s)$ may be obtained from 
$\langle \ee^{-sK} \rangle_0 \simeq \ee^{-N\tobs \psi_K(s)}$~\cite{psi-foot}.
%
We will consider
averages of one-time quantities (like $\langle n_i(t) \rangle_s$), as well as quantities that 
depend on the whole trajectory (like $\langle K \rangle_s$).
Averages of one-time quantities are independent of the time $t$ at which they are evaluated, except for
initial and final `transient' regimes near $t=0$ and $t=\tobs$.  Unless otherwise stated, we evaluate
all one-time quantities at time $t=\tobs/2$, which is representative of the steady-state regime.

As discussed in~\cite{kcm-transition},
properties of $s$-ensembles in the East model may be obtained by analysis of the operator
\begin{equation}
\WW_K(s) =  \sum_i \hat{n}_{i-1} \big[  \ee^{-s} (1-c)\sig^-_i + \ee^{-s} c \sig^+_i  - (1-2c)\hat{n}_i -c   \big]  .
\end{equation}
The largest eigenvalue of $\WW_K(s)$ is equal to $-N\psi_K(s)$.  Let the left and right eigenvectors corresponding
to this eigenvalue have elements
$u_\CC$ and $v_\CC$ respectively, normalised such that $\sum_\CC u_\CC v_\CC=1$.
Then the steady-state probability of configuration $\CC$ within the $s$-ensemble 
is $p_\CC = u_\CC v_\CC$.

The large deviations of $K$ are closely related to those of the `time-integrated escape rate'~\cite{kcm-transition},
\begin{equation}
R[\CC(t)] = \int_0^{\tobs}\!\!\mathrm{d}t \, r(\CC(t)) ,
\end{equation}
where $r(\CC)$ was defined in (\ref{equ:def-r}), above.
By analogy with (\ref{equ:s-ens}), we define a `$\nu$-ensemble' by
\begin{equation}
\mathrm{Prob}[\CC(t),\nu] = \mathrm{Prob}[\CC(t),0] \cdot \frac{ \ee^{\nu R} }{ \langle \ee^{\nu R} \rangle_0 }  .
\label{equ:nu-ens}
\end{equation}
The properties of this ensemble may be obtained from the operator
\begin{equation}
  \WW_R(\nu) = \ee^{s} \WW_K(s), \qquad \ee^{s} = 1 - \nu  .
  \label{equ:nu-s}
\end{equation}
The largest eigenvalue of $\WW_R(\nu)$ is therefore $-N\psi_R(\nu)$ with $\psi_R(\nu) = \ee^s \psi_K(s)|_{\ee^s=1-\nu}$, and the eigenvectors associated
with this eigenvalue are the same $u_\CC$ and $v_\CC$ found by diagonalising $\WW_K(s)$.   We use $\langle \cdot \rangle_\nu$
to represent averages with respect to $\mathrm{Prob}[\CC(t),\nu]$.  From (\ref{equ:nu-s}), it follows that the $\nu$-ensemble
and the $s$-ensemble contain the same information (at least for $\nu<1$). 

Since the dynamics of the East model obeys detailed balance, and the activity $K$ is time-reversal symmetric,
the operator $\WW_K(s)$ may be symmetrised~\cite{kcm-transition} as 
$\HH_K(s) = \ee^{\beta\sum_i \hat{n}_i/2} \WW_K(s) \ee^{-\beta\sum_i \hat{n}_i/2}$; the same transformation also symmetrises
$\WW_R(\nu)$.
It follows~\cite{Jack-Sollich-PTP}  that both $s$-ensembles and $\nu$-ensembles 
may be described by ``auxiliary Markov models'' that obey detailed balance
with respect to the distribution $P_s(\CC) \propto u_\CC^2 \ee^{-\beta\sum_i n_i}$.   
Hence, we write
\begin{equation}
u_\CC = \ee^{-\Delta V_\CC/2}
\label{equ:def-VC}
\end{equation}
and interpret $\Delta V_\CC$ as an `effective potential' that acts to drive the system into the 
$\nu$-ensemble (see also~\cite{Simon09}).   
If $\Delta V_\CC$ is
known then steady state averages of one-time quantities may be obtained by the methods of equilibrium statistical mechanics,
using the energy function $\beta E(\CC) = \beta \sum_i n_i + \Delta V_\CC$.  The main purpose of this paper is to investigate
the `effective potential' $\Delta V_\CC$ that describes the $\nu$-ensemble for the East model.

\subsection{Observables and correlation functions}
\label{sec:def-correl}

%
%

As well as the effective potential $\Delta V_{\CC}$, which fully describes the biased state,
we also consider several simpler observables that provide
insight into the response to the bias.  These include the mean escape rate, 
\begin{equation}
r(\nu) = \frac{1}{N\tobs}\langle R \rangle_\nu ,
\end{equation}
 which
 indicates how strong
 the bias $\nu$ must be, in order to produce a particular deviation of $R$ from its average.  From the definition
of the dynamical free energy $\psi_R(\nu)$ as the long-time limit of $-1/(N\tobs)\ln\langle\ee^{\nu R}\rangle_0$ it follows that $r(\nu)=-\psi_R'(\nu)$.
The derivative $\chi_R(\nu)=r'(\nu)=-\psi_R''(\nu)$,
then indicates the size of the fluctuations of $R$, via
\begin{eqnarray}
\chi_R(\nu) & = \frac{1}{N\tobs} [ \langle R^2 \rangle_\nu - \langle R \rangle^2_\nu] 
 \nonumber \\ &
= \frac{1}{N\tobs} \sum_{ij} \int_0^{\tobs}\!\! \dd t\, \dd t' \langle \delta r_i(t) \, \delta r_j(t') \rangle_\nu  ,
\label{equ:chi-rr}
\end{eqnarray}
where 
\begin{equation}
r_i=(1-c) n_{i-1} n_i + c n_{i-1} \nbar_i
\label{equ:def-ri}
\end{equation}
is the escape rate at site $i$. We use the notation $\nbar_i = 1-n_i$
and $\delta O = O - \langle O \rangle$.

To characterise spatial correlations in the biased state, we define
\begin{equation}
C(x) = \langle \delta n_i\, \delta n_{i+x} \rangle_\nu .
\label{equ:cx}
\end{equation}
At equilibrium, $C(x)=c(1-c)\delta_{x,0}$, since the energy function of the system lacks any interactions between spins.
However, the presence of a non-zero effective potential $\Delta V_{\CC}$ modifies $C(x)$, which then provides a simple
characterisation of response to the bias.

Finally, an observable that is very useful for probing the structure of biased states is a probability distribution for domain sizes,
denoted by $p(d)$.
Here, each domain consists of a single up-spin and all the down-spins to its right, up to (but not including) the next up-spin. 
This definition is slightly different from the natural definition of domains
in (for example) a one-dimensional Ising model, where a block of adjacent up spins would form a single domain:  in our case, each
up spin forms its own separate domain.  Our definition is motivated by the fact that
up spins are typically quite rare, and the spacing between these spins is a dominant factor in determining dynamical behaviour.
More formally, we define 
observables $U_{i,i+r}$ and $D_{i,i+r}$ which are equal to unity if spins $i$ to $i+r$ are all
up ($U$), or all down ($D$), and zero otherwise.  That is,
\begin{eqnarray}
U_{i,i+r} &= \prod_{j=i}^{i+r} n_j , \qquad
D_{i,i+r} &= \prod_{j=i}^{i+r} \nbar_j  .
\label{equ:UD}
\end{eqnarray}
In this notation, the domain size distribution is
\begin{equation}
  p(d) = \frac{ \langle n_i \cdot D_{i+1,i+d-1} \cdot n_{i+d} \rangle_\nu}{ \langle n_i \rangle_\nu} .
\label{equ:def-pd}
\end{equation}
At equilibrium, one has the distribution $p^0(d) = c(1-c)^{d-1}$. 


\subsection{Range of effective interactions}

For a finite system, it is always possible to write the eigenvector $u_\CC$ as $\ee^{-\Delta V_\CC/2}$,
as in (\ref{equ:def-VC}).  However, the question of whether $\Delta V_\CC$ has the features
expected of a ``physical'' potential is a subtle one.  For spin models, Ising-like potentials
such as $V_\CC = \mu \sum_i n_i - \epsilon \sum_i n_i n_{i+1}$ are familiar,   
but $\Delta V_\CC$ may also contain two-body interaction terms like $n_i n_{i+x}$ of larger range $x$, or many-body terms like $n_i n_{i+1}n_{i+2}$, which is a three-body interaction of range 2. While these are less familiar in physical situations, we show below that $\Delta V_\CC$ for the East model with $\nu>0$ must contain a combination of such terms with unbounded range (up to the system size), and we argue that this behaviour can be expected more generally. 
Specifically, if the longest range interaction term in $\Delta V_\CC$ has
range $B-1$ then the domain size distribution $p(d)$ must decay exponentially for $d>B$ (a derivation is given in~\ref{app:pd-exp}).
In the following, we will present numerical and analytic evidence that for $\nu>0$, $p(d)$ decays faster than exponentially
at large $d$, indicating the effective potential contains \emph{long-ranged interactions}:
that is, no effective potential with interactions of bounded range can fully
represent $\Delta V_\CC$ as the system size $N\to\infty$.  In this sense, the states that occur for $\nu>0$
in these systems are qualitatively more complex than those of classical spin models at equilibrium.  This is part
of the reason for the rich behaviour that has been observed in the large deviations of such 
systems, even in one dimension~\cite{JL,Bod-Der,Imparato09,kcm-transition}.

We also note in passing that effective potentials similar to $\Delta V_\CC$ have been considered in the mathematical
literature.  For example, instead of considering configurations $\CC$ within long trajectories, one can
consider the configuration of a single (two-dimensional) lattice plane, within a three-dimensional Ising 
model~\cite{Maes99-ising}.  Similar questions also arise within the renormalisation group~\cite{Sokal93}.
The main question that has been addressed there is whether the effective
interactions are ``Gibbsian'': that is, whether $\Delta V_\CC$ is a ``reasonable Hamiltonian'' in the sense
defined rigorously in~\cite{Sokal93}.  (The idea is based on considerations of locality, in the sense that the interactions encoded by
$\Delta V_\CC$ should decay sufficiently quickly with interaction range~\cite{Sokal93}.)  In the following, 
we focus on the specific interactions that we find within the East model: whether these interactions are Gibbsian or not is
a question that we postpone for later studies.

\section{Overview of response to bias, and numerical results}
\label{sec:num}

This Section introduces the most important features of the response of the East model to the bias $\nu$.
We present numerical results that illustrate the structure that develops when the bias $\nu$ is applied, and we
also discuss two concepts that are important for interpreting this structure: the notion
of `quasiequilibrium', and the existence of scaling behaviour at small $c$.

\subsection{Mean activity and susceptibility}

The $\nu$-ensemble may be sampled numerically using transition path sampling (TPS)~\cite{TPS-review}: 
we follow the methods
described in~\cite{hedges09,elmatad10}.  For small systems (at least up to $N=14$)
one may also diagonalise the operator $\WW_R(\nu)$ exactly, and obtain $\Delta V_\CC$ directly
from its eigenvectors.  (It is convenient
to first symmetrise the operator, as described in Section~\ref{sec:model}.)


\begin{figure}
\hfill
\includegraphics[width=7cm]{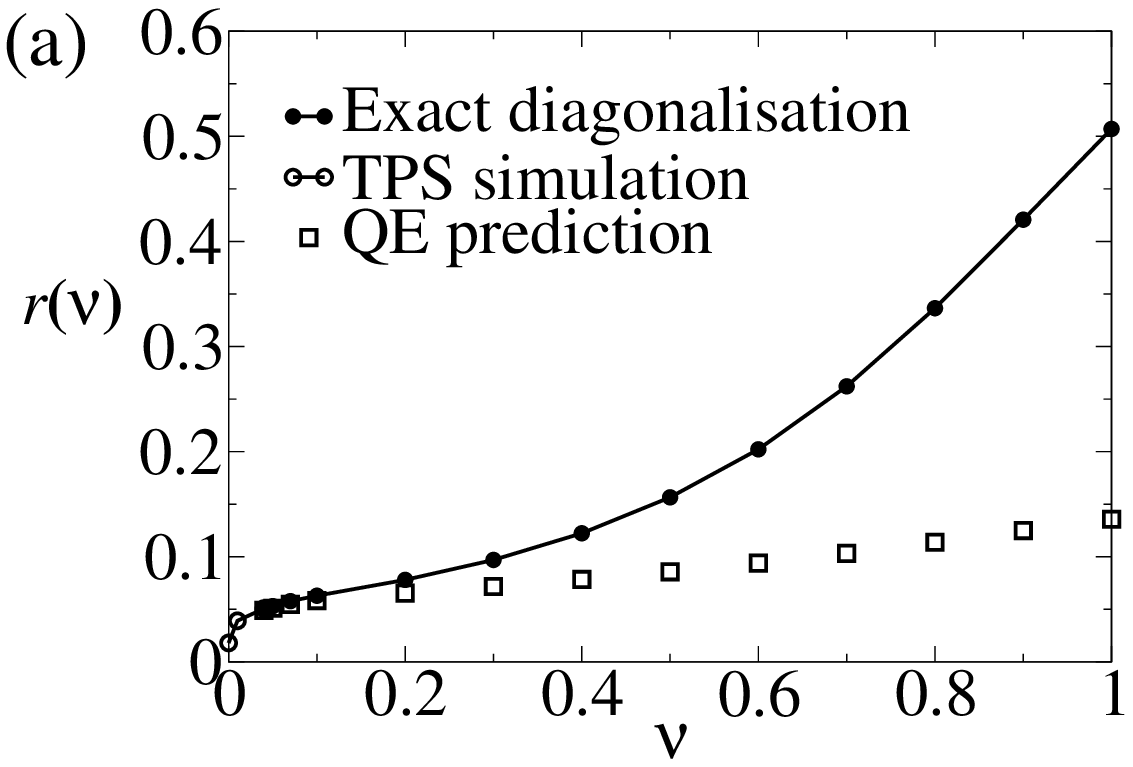}  
\includegraphics[width=7cm]{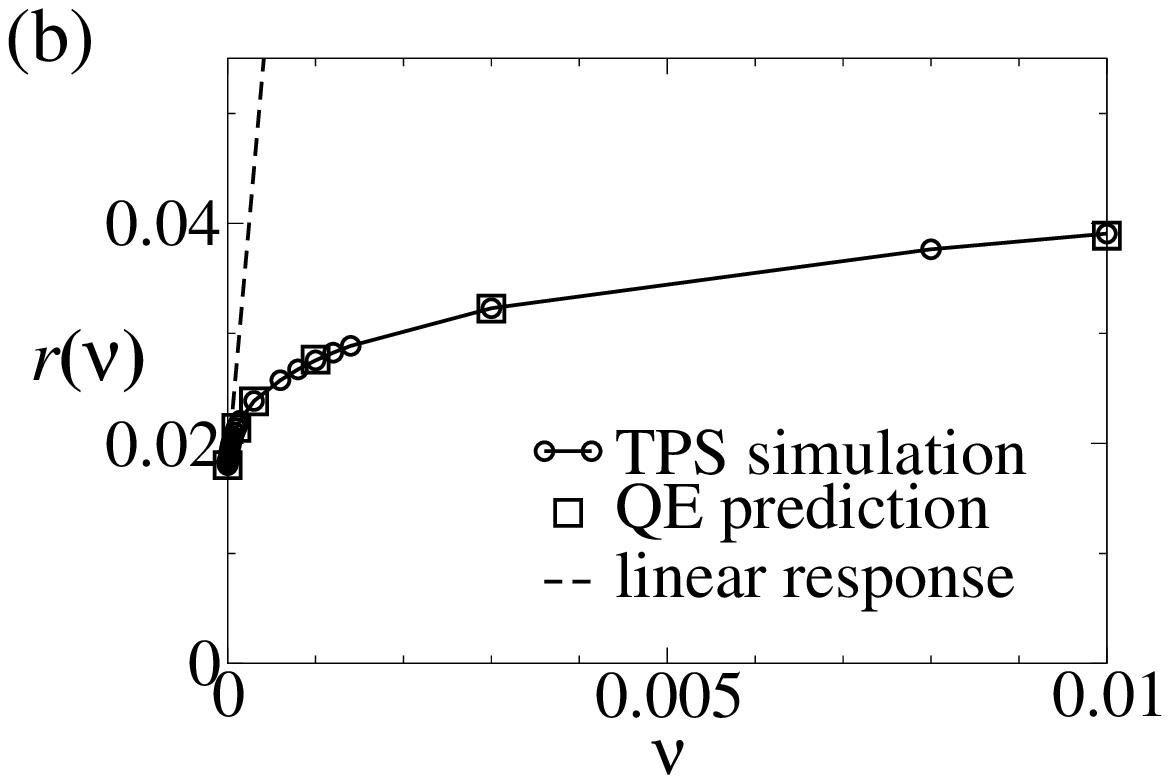} 
\par\hfill
\includegraphics[width=7cm]{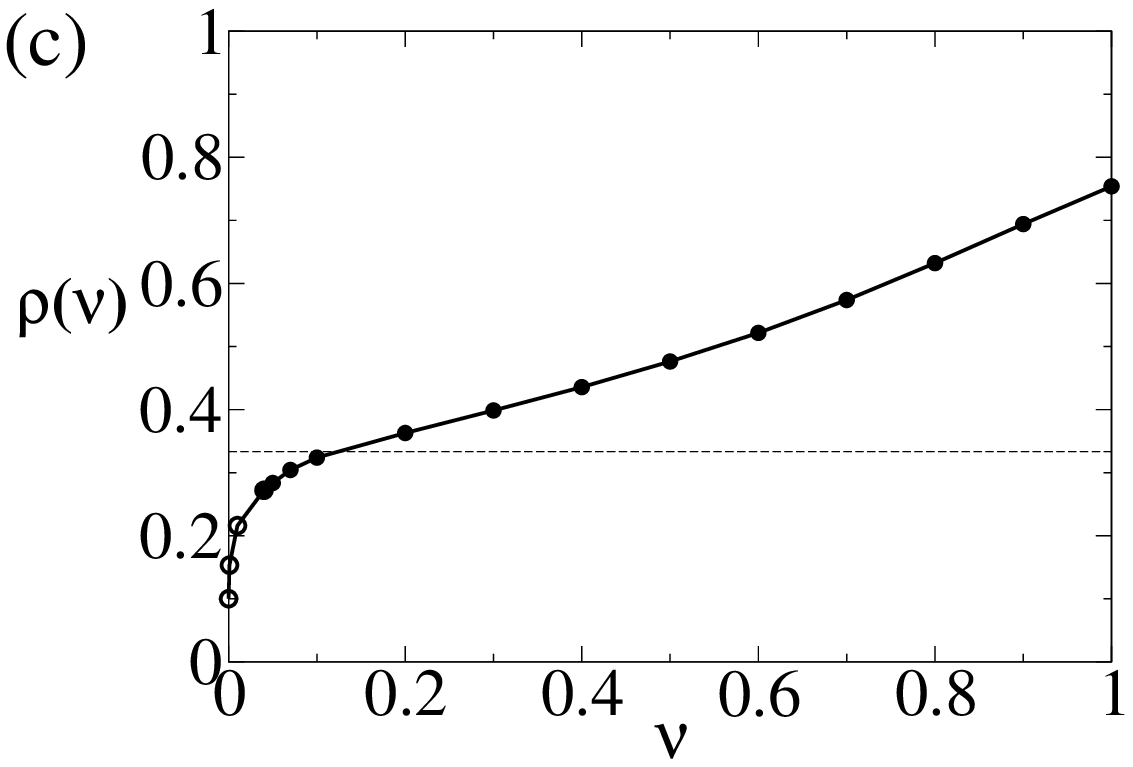} 
\includegraphics[width=7cm]{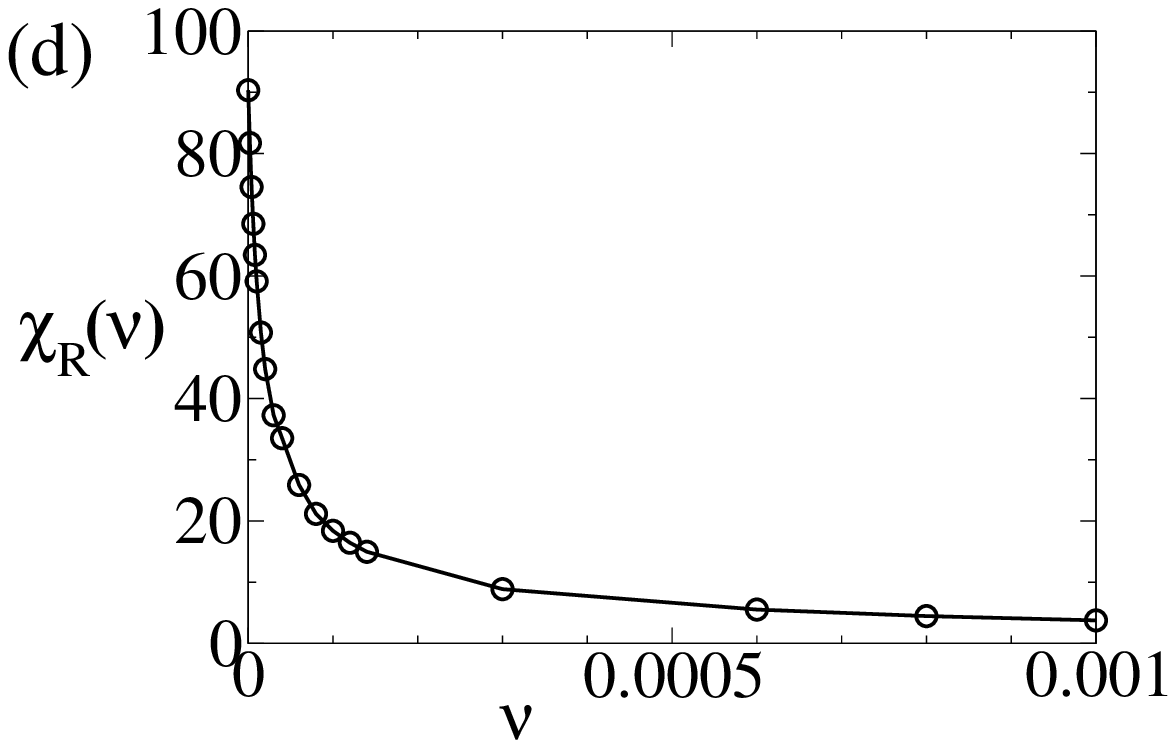} 
\caption{
	The mean escape rate $r(\nu)$, up-spin density $\rho(\nu)=\langle n_i\rangle_\nu$, and 
	the susceptibility $\chi_R(\nu)$, in the East model for $c=0.1$. 
	(a)~Mean escape rate $r(\nu)$ for the range $0<\nu<1$ (corresponding to $-\infty<s<0$ in the $s$-ensemble). This is
	compared with the prediction of the quasiequilibrium (QE) assumption (\ref{equ:qe-pred}).
	(b)~Mean escape rate $r(\nu)$ for small $\nu$. The QE assumption holds quite accurately. 
	(c)~Mean up-spin density $\rho(\nu)$. The dashed line indicates $\rho(\nu)=\frac13$, which is the predicted
	density in the limit $c\ll\nu\ll 1$ (see text).  
	(d)~The susceptibility $\chi_R(\nu)$, for small $\nu$. 
	In all panels, solid circles show data from exact diagonalisation of the operator 
	$\WW_R(\nu)$ and open circles are data from transition path sampling (TPS)~\cite{TPS-review}.  
	The system sizes are $N=14$ (exact diagonalisation), $N=32$ (TPS for $\nu\geq0.01$) and $N=64$ (TPS for $\nu<0.01$).  
	%
	%
}
\label{fig:rnu}
\end{figure}

Fig.~\ref{fig:rnu} shows numerical results for the average escape rate in the $\nu$-ensemble, and the associated
susceptibility $\chi_R(\nu)$. 
	For the exact diagonalisation, we show results for $N=14$: we also analyzed the case $N=12$
	for which the results agree to within the symbol sizes, indicating
	that finite-size effects are small.
	In the TPS calculations, we vary $\tobs$ according to the state point, to ensure
    convergence of the large-$\tobs$ limit.  We have not carried out a detailed analysis of finite size effects
    but we do ensure that systems are significantly larger than almost all domains that appear
    at each state point (see discussion in future sections).  
    We therefore expect finite size effects to be small in these cases too.
    
 In Fig.~\ref{fig:rnu}(d), it is striking that 
$\chi_R(\nu)$ is large as $\nu\to0$, so that the average escape rate $r(\nu)$ responds
strongly to the bias $\nu$.  We emphasise however that this susceptibility
is finite even as $N,\tobs\to\infty$.  To show this, we use a property of the East model at equilibrium. Suppose that $A$
and $B$ are two observables that depend on the spins only in non-overlapping regions of the system. By this we mean that
all the spins that $A$ depends on are to the left of those in $B$, or vice versa. Then one has
\begin{equation}
\langle \delta A(t) \delta B(t')\rangle_0=0  .
\label{equ:ABzero}
\end{equation}
This result~\cite{Ritort-Sollich} follows from the directionality of the kinetic constraint in the East model, which means that
information can only flow from left to right in the system. Causality then implies that the state of 
spins with $j>i$ at time $t$ cannot affect the
behaviour of spin $i$ for any $t'>t$.  In fact, this property holds both at
equilibrium and during relaxation towards equilibrium, which is sufficient to prove that a restricted version of (\ref{equ:ABzero}) 
holds both in and out of equilibrium.  That is, (\ref{equ:ABzero}) holds both in equilibrium and out of equilibrium
as long as $t'>t$ and observable $B$ is localised to the left of observable $A$.  At equilibrium, time-reversal
symmetry then implies that (\ref{equ:ABzero}) holds for all $t$ and $t'$. 

Combining Eqs (\ref{equ:def-ri},\ref{equ:ABzero}), one finds that
$\langle \delta r_i(t) \,\delta r_j(t')\rangle_0 = 0$ for $|i-j|>1$.  And for any $i,j$ with $|i-j|\leq 1$, the finite
spectral gap~\cite{East-gap} of the operator $\WW$ means that $\langle \delta r_i(t) \,\delta r_j(t')\rangle_0$
decays exponentially for long times.  Thus, the combined integral and sum in (\ref{equ:chi-rr}) leads to
a finite result, at $\nu=0$.
We also note that the ratio
$r(0)/\chi_R(0)$ sets a natural scale for $\nu$: the strength of bias required to introduce an $O(1)$ relative change in activity.
From (\ref{equ:chi-rr}), this can be estimated to be is of the order of $\tau_0^{-1}$ where $\tau_0$ is the equilibrium relaxation time, of the order of the 
inverse spectral gap.
In fact, the bias $\nu$ has a hierarchy of natural scales, of which this is just the smallest.  We return to this point in later Sections.


\subsection{Quasiequilibrium condition}
\label{sec:qe-simple}

One effect of the parameter $\nu$ is to bias the system away from its equilibrium state.
However, an important observation for interpreting the results of this article is that
some degrees of freedom in the East model remain `quasiequilibrated' in the presence of the bias, 
at least as long as $c$ is small and $\nu$ is not too large.
This means that even if configurations of the system have probabilities far from their equilibrium values,
\emph{ratios} of probabilities for some configurations may be almost unaffected by the bias.  In other words, one
may identify pairs of 
states for which $\Delta V_\CC - \Delta V_{\CC'}$ is small, even if the absolute values of the effective potential are large.

To illustrate this situation,
suppose that spin $i$ is ``facilitated'' in the East model (that is, $n_{i-1}=1$).  Then spin $i$ will
flip on a relatively rapid time scale of order $c^{-1}$.  At low temperatures (small $c$), it is likely that spin $i-1$ will flip only 
on a much slower time scale.  In this case, spin $i$ typically flips many times before spin $i-1$ flips at all.  
Holding all other spins in the system constant, one then compares configuration $\CC$ (where $n_i=0$), with 
configuration $\CC'$ (where $n_i=1$).  Regardless of whether the system was initially in $\CC$ or $\CC'$, the rapid flips
of spin $i$ mean that
the ratio of probabilities of these configurations after a time of order $1/c$ will be very close to $c/(1-c)$, which
is equal to their ratio at equilibrium. (Here we exploit the smallness of $\nu$ to neglect the effects of
the bias on local spin flips.) It follows that
$\langle n_{i-1} n_{i} \rangle_\nu \approx \langle n_{i-1} \rangle_\nu c$.  To the extent that this holds, 
one has from (\ref{equ:def-ri}) that
\begin{equation}
\langle r_i\rangle_\nu \approx 2c(1-c)\langle n_i \rangle_\nu, \qquad
r(\nu) \approx 2c(1-c) \langle \rho\rangle_\nu
\label{equ:qe-pred}
\end{equation}
where $\rho = N^{-1}\sum_i n_i$. These relations indicate that the escape rate and the density
of up spins in the biased ensemble are tightly correlated for the East model, as found numerically
in~\cite{Merolle,Jack06}

Fig.~\ref{fig:rnu} confirms  that the quasiequilibrium relation (\ref{equ:qe-pred}) does hold
quite accurately for $\nu\lesssim c$. For larger $\nu$, quasiequilibrium 
breaks down: the response for $\nu=O(1)$ changes in character, resulting in a steady increase in the escape rate but departure
from quasiequilibrium.  For $c \ll \nu \ll 1$, we expect a plateau in $r(\nu)$, which represents a quasiequilbrium
state with $\langle n_i \rangle_\nu \approx \frac13$ (see Sec.~\ref{sec:hier} below).  
The data in Fig.~\ref{fig:rnu}(c) are consistent with such a regime, although numerical limitations prevent us from obtaining 
results at smaller $c$, which would allow this hypothesis to be tested further.
In the regime $\nu>1$, the function $r(\nu)$ increases smoothly, 
finally saturating in the state with all up spins as $\nu\to\infty$ (data not shown).

The quasi-equilibrium argument above can also be applied to individual
configurations, not just averages over configurations.  It then says
that while a spin at site $i-1$ is up ($n_{i-1}=1$), it causes the escape
rate of the configuration $R=\sum_j r_j$ (or more precisely its
contribution from site $i$) to be higher by $2c(1-c)-2c^2(1-c) =
2c(1-c)^2 \approx 2c$ than the equilibrium average of $2c^2(1-c)$ per
site.  Conversely, a down-spin lowers the escape rate, changing it by
$-2c^2(1-c)\approx -2c^2$.

\subsection{Spatial correlations}
\label{sec:cx_pd}

We now turn to spatial structure in the $\nu$-ensemble.
Our overall aim is to arrive at a representation of the effective potential $\Delta V_\CC$,
but we begin
by considering simple measures of order, to gain an overview of the structure of the system.

\begin{figure}
\hfill
\includegraphics[width=5cm]{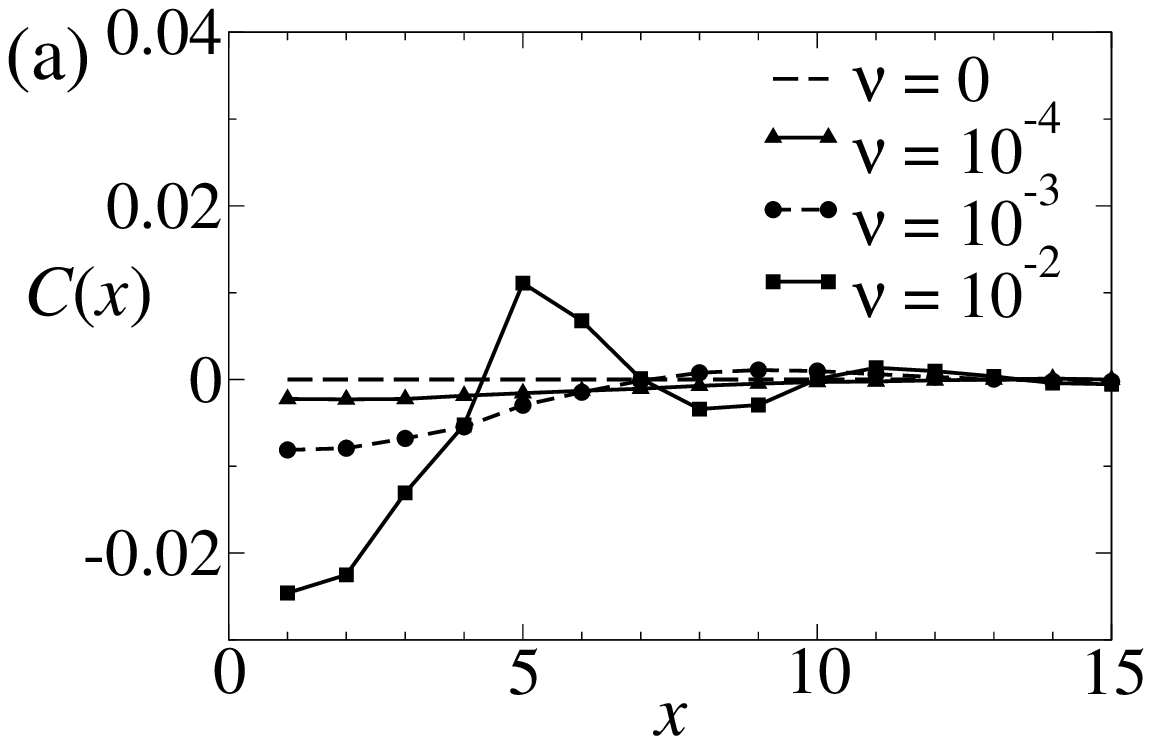} 
\includegraphics[width=5cm]{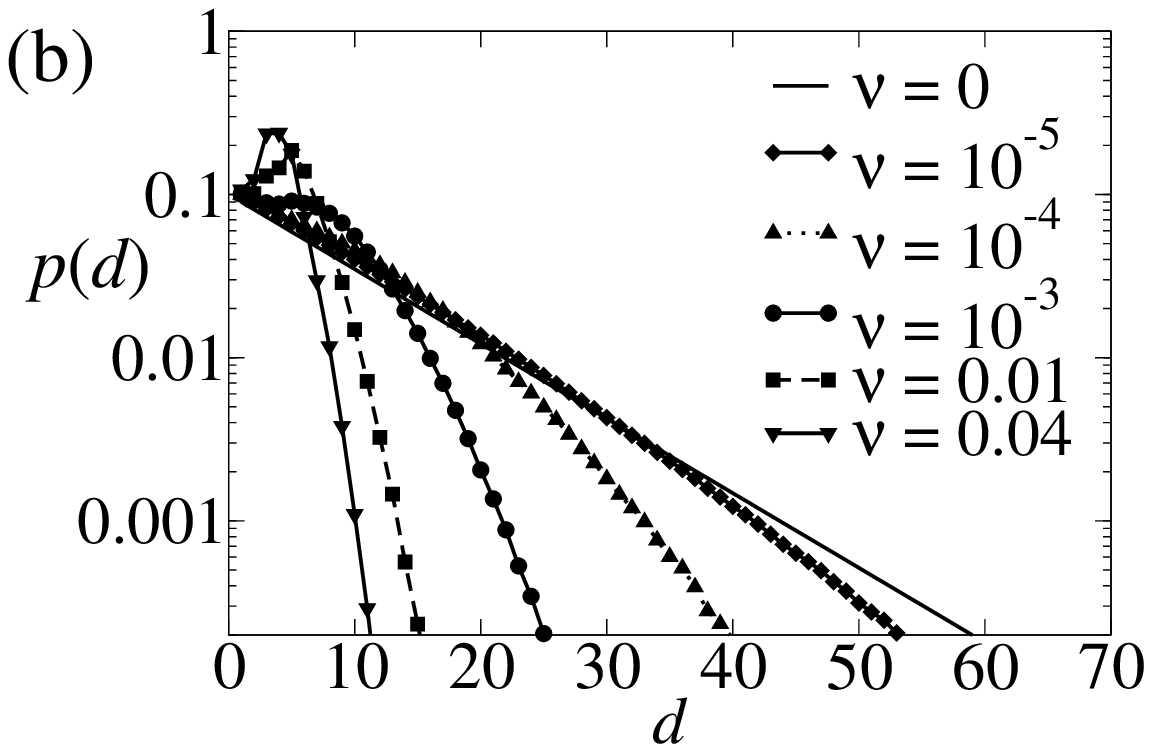}
\includegraphics[width=5cm]{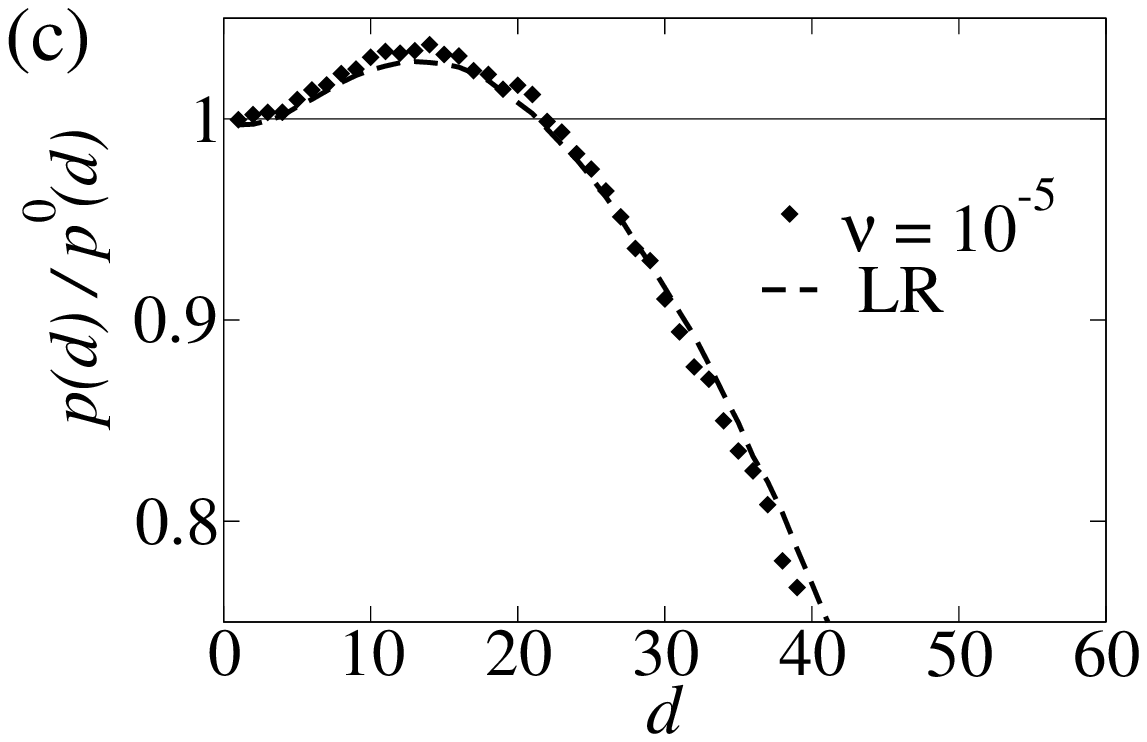}
\caption{
  Spatial structure in the East model for $c=0.1$.  (a)~Correlation function $C(x)$.
  (b)~Distribution of domain sizes $p(d)$.
  (c)~Plot of $p(d)/p^0(d)$ for $\nu=10^{-5}$, where $p^0(d)$ is the equilibrium distribution of domain sizes.
  The dashed line shows the prediction of linear response theory (LR),  
  obtained via the fluctuation formula (\ref{equ:pt1}), by numerical evaluation of the relevant equilibrium correlation function.
}
\label{fig:gr-pd}
\end{figure}

In Fig.~\ref{fig:gr-pd}(a), we show the two-point correlation function 
$C(x)$ defined in (\ref{equ:cx}).
For $\nu>0$, up spins appear to `repel' each other: the probability of finding two up-spins close together
is suppressed, with respect to the value for independently fluctuating spins found at equilibrium.  
For $\nu\gtrsim 10^{-3}$, one observes oscillations in
$C(x)$: just beyond the range of the short-ranged `repulsive' correlations is a `nearest neighbour peak' where
up spins are more likely to be found.  

The correlations that are apparent in $C(x)$ remain quite weak as $\nu$ increases, 
in stark contrast to the large changes in $r(\nu)$ and $\chi_R(\nu)$ shown in Fig.~\ref{fig:rnu}.  
A more revealing measurement of the system's structure in the $\nu$-ensemble is to consider the distribution
of domains of down spins $p(d)$ as defined in (\ref{equ:def-pd}).
Figs.~\ref{fig:gr-pd}(b,c) show that there is considerable
structure in the tails of $p(d)$.  
From a physical point of view, the key aspects of $p(d)$ are that the distribution narrows as $\nu$ increases, with large domains
being strongly suppressed, while small domains (for example $d=1$ and $d=2$) are only weakly affected.    The probability associated
with the larger domains is transferred to intermediate domain lengths which become increasingly common, eventually leading to a peak
in $p(d)$ at an emergent length scale $d^*>1$.  Since the relaxation times for the largest domains are longest, 
then (\ref{equ:chi-rr}) indicates that
the suppression of large domains is directly linked with the suppression of the susceptibility $\chi_R(\nu)$ 
as $\nu$ is increased [recall Fig.~\ref{fig:rnu} and Eq.~(\ref{equ:chi-rr})].
We also note
in passing that the arguments leading to the 
quasiequilibrium condition (\ref{equ:qe-pred}) predict $p(d=1)=c$, which holds quite accurately in Fig.~\ref{fig:gr-pd}(b).
In the following Sections, we will concentrate on $p(d)$ as an observable
that reveals the dominant correlations within the biased steady state.

\subsection{Scaling behaviour}

\begin{figure}
\hspace{2cm} \includegraphics[width=8cm]{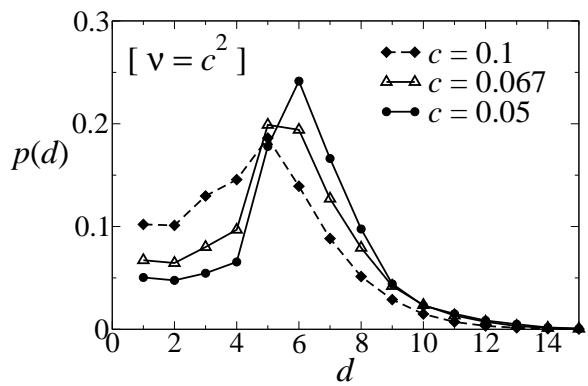}  
\caption{Domain distribution $p(d)$, for $\nu=c^2$ and varying $c$.  This distribution is peaked
around an emergent length scale and the peak becomes sharper as $c$ is reduced, indicating that
this length scale can be associated with the limit of small-$c$, given $\nu=c^2$.
In this limit, scaling arguments (see Sec.~\ref{sec:hier} below) predict that
$p(d) = O(c)$ for $d\leq 4$ while $p(d)$ has a positive limit for $d\geq 5$.  
%
%
This is consistent with the data, although the system is still quite far from the small-$c$ limit.
}
\label{fig:sc2}
\end{figure}


It is well-known that length and time scales are intrinsically connected in the East model,
both at equilibrium and in the aging behaviour~\cite{Sollich-Evans}.  Indeed, it is clear from (\ref{equ:tau-ell})
that the model obeys scaling relations in the limit $c\to0$, where both length and time scales diverge.
In the presence of the bias $\nu$, it will be useful to consider limits where both $c$ and $\nu$ go to zero
together: we typically fix $\nu=c^b$ and then take $c\to0$.  Fig.~\ref{fig:sc2} shows numerical results
with $\nu=c^2$, as $c$ is varied.  One sees that $p(d)$ develops a peak at an emergent length scale, and
that this peak becomes increasingly well-defined as $c$ is reduced.  In the next Section,
use a perturbative scheme to analyse this kind of behaviour in more detail. This leads to a physical
picture that we outline in Section.~\ref{sec:hier}, where we identify the limit of $c\to0$ at $\nu=c^b$ with
a sharply-defined $b$-dependent length scale.

\section{Linear response theory}
\label{sec:pert}

At first order in $\nu$, we can obtain a formula for the effective potential $\Delta V_\CC$ using linear response theory (perturbation
theory about the equilibrium state).
Consider a one-time quantity $f$, and its average $\langle f \rangle_\nu$.
For small $\nu$, Eq.~(\ref{equ:nu-ens}) gives
\begin{equation}
\langle f \rangle_\nu=\langle f \rangle_{0} + \nu \langle \delta f  \,
\delta R \rangle_0 + O(\nu^2)  .
\label{equ:LR}
\end{equation}
We note that $\langle \delta f  \,\delta R \rangle_0 = \langle f  \,\delta R \rangle_0$: it is
sometimes convenient to use this latter form in the following.
In addition, the first-order term in (\ref{equ:LR})
may be written as
\begin{equation}
\delta \langle f \rangle = \nu \sum_j \int_0^{\tobs} \!dt\, \langle \delta f(t')\, \delta r_j(t) \rangle_0.
\label{equ:pt1}
\end{equation}
where we used (\ref{equ:def-ri}), and we evaluate $f$ at $t'=\tobs/2$ to avoid transient regimes, as discussed above.
For large enough $\tobs$ (and using time-reversal symmetry at equilibrium), 
this may also be written as~\cite{kcm-transition}
\begin{equation}
  \delta \langle f \rangle = 2\nu \sum_j \int_0^{\infty} \!dt\, \langle \delta f(0) \, \delta r_j(t) \rangle_0.
\label{equ:pt_prop}
\end{equation}

To obtain $\Delta V_\CC$, we take $f=e_\CC$ to be an indicator function, which has a value of unity if the system
is in configuration $\CC$, and zero otherwise.   One finds
\begin{eqnarray}
 \Delta V_\CC & =  -\frac{\delta \langle e_{\CC} \rangle}{ \langle e_{\CC}\rangle_0 } + O(\nu^2) 
 \nonumber \\
 & = - 2\nu \sum_j \int_0^{\infty} \!dt\, \frac{ \langle \delta r_j(t) e_{\CC}(0) \rangle_0 }{ \langle e_{\CC}\rangle_0 }
     + O(\nu^2)  .
\label{equ:eCC}
\end{eqnarray}
We define 
\begin{equation}
R_\CC \equiv \frac{1}{\langle e_{\CC}\rangle_0 } \sum_j \int_0^{\infty} \!dt\, \langle \delta r_j(t) e_{\CC}(0) \rangle_0
\end{equation}
as the propensity~\cite{propensity}
for activity $R$ associated with configuration $\CC$: this may
be obtained by averaging the observable $R$ over trajectories with initial condition $\CC$. 
Hence, the effective potential 
of configuration $\CC$ in the presence of the bias $\nu$ is given, to leading order, by its propensity:
\begin{equation}
\Delta V_\CC = -2\nu R_{\CC} + O(\nu^2) .
\label{equ:dV-pert}
\end{equation}
Hence, at this order, all effective interactions may be obtained numerically by the relatively simple
procedure of calculating propensities.  However, this does not address the central challenge associated 
with characterising the effective interactions.  After all, if the system size is $N$, then the  propensity 
$R_\CC$ is a set of $2^N$ numbers: one must still address how the dependence of $R_\CC$ on the structure
of the system can be represented in a useful way.  This will be the main strength of the variational approach in Section~\ref{sec:var}, below.

\subsection{Enhancement of the density $\langle n_i\rangle_\nu$}
\label{sec:pert-rho}

In the remainder of this section,
we use (\ref{equ:pt1}) to investigate the effect of a small bias $\nu$ on the structure of the system.
We begin with the response of the mean density of up-spins:
$\delta \langle n_i\rangle$.  
We recall from  (\ref{equ:ABzero}) that equilibrium two-time correlation functions vanish unless they involve observables on overlapping regions
of the chain, so that
\begin{equation}
 \delta \langle n_i \rangle = 2\nu  \int_0^\infty\!\mathrm{d}t\, \langle \delta n_i(0) [ \delta r_{i+1}(t) + \delta r_{i}(t) ] \rangle_0 .
 \label{equ:n-pert}
\end{equation}
The dominant contribution to this response appears because $r_{i+1}>0$ if and only if $n_i=1$: physically, this is the statement
that if spin $i$ is up then spin $i+1$ is able to flip.
The correlation function in the expression for $\delta \langle n_i \rangle$ that corresponds to this effect is
\begin{eqnarray}
\langle \delta n_i(0) \delta r_{i+1}(t) \rangle_0 &  = \langle \delta n_i(0) [ (1-c)n_i(t) n_{i+1}(t) + c n_i(t) \nbar_{i+1}(t)] \rangle_0
\nonumber \\ & \approx 2c(1-c) G_1^0(t) ,
\label{equ:nr}
\end{eqnarray}
where the single-site autocorrelation function $G_1(t)=\langle \delta n_i(t')\, \delta n_{i}(t'+t) \rangle_\nu$ 
and the superscript $0$ indicates
that we evaluate it at equilibrium  ($\nu=0$). The  approximate equality holds if 
$\langle n_i(0) n_i(t) n_{i+1}(t)\rangle \approx \langle n_i(0) n_i(t) \rangle \langle n_{i+1}(t)\rangle$, which is to be expected,
 based on quasiequilibrium
arguments along the lines of those in Section~\ref{sec:qe-simple}.

The function $G_1^0(t)$ 
decays from $c(1-c)$ to 0 on the time scale $\tau_0$.  We approximate the time-integral in Eq.~\ref{equ:n-pert} by multiplying
its maximal value by this relaxation time, leading to  
$\delta \langle n_i \rangle \simeq 4\nu c^2\tau_0$,
which diverges as $c\to0$ (recall that $\tau_0$ diverges faster than any power of $1/c$).  
Thus, the response of the density $\rho$ to the field $\nu$ diverges very quickly as $c\to0$.  The strong quasiequilibrium
correlation between $\rho$
and $R$ means that $R$ also responds very strongly, consistent with the large value of $\chi_R(0)$ shown in Fig.~\ref{fig:rnu}.

\subsection{Spatial structure in the biased state}
\label{sec:rel_enh}

Given that the
 up-spin density 
$\langle n_i\rangle$ responds strongly to the bias $\nu$ already within the linear
response regime, the next step is to consider the spatial correlations that accompany this increase.
%
%

\begin{figure}
\hspace{2.5cm}\includegraphics[width=7cm]{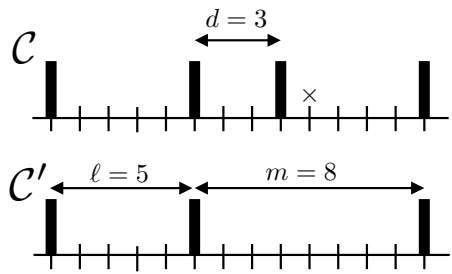}
\caption{
  Sketch illustrating the configurations $\CC$ and $\CC'$ discussed in Section~\ref{sec:rel_enh}.
  Black bars indicate up spins ($n_i=1$). 
  At low temperatures, the hierarchy of energy scales in the East model means that $\CC$ will
  almost certainly relax to $\CC'$ on a time scale $\tau_3$ given by (\ref{equ:tau-ell}); this process
  is much faster than any other local relaxation mechanism.  The propensity $R_\CC$ differs
  from $R_{\CC'}$ primarily due to the facilitated site that is marked with a $\times$.
} \label{fig:d-sketch}
\end{figure}

We first 
consider two configurations $\CC$ and $\CC'$, with $\CC$ having one more up spin than $\CC'$, as shown in Fig.~\ref{fig:d-sketch}.  
On increasing $\nu$, we expect the probability
of $\CC$ to be enhanced with respect to $\CC'$, since the density of up spins is increasing.  
To obtain the enhancement of $\CC$ with respect to $\CC'$, it is sufficient to
estimate
the propensity difference $R_\CC - R_{\CC'}$.  We focus on the domains in the system, defined as in Sec.~\ref{sec:cx_pd} (each
up spin starts a new domain, and the domain length is the distance to the next up spin to the right).
  As in Fig.~\ref{fig:d-sketch}, 
  let the domains in the region of interest of $\CC'$ be $(\dots,\ell,m,\dots)$ and those in $\CC$ be $(\dots,\ell,d,m-d,\dots)$.
  The two configurations coincide exactly in regions indicated by $(\dots)$.

We further assume that
$\alpha_m, \alpha_\ell > \alpha_d$, where the barrier height $\alpha_\ell$ was defined in (\ref{equ:alpha}).
Then, the hierarchical relaxation in the East model means that as
$c\to0$, the local time evolution of $\CC$ and $\CC'$ is deterministic, in the sense that
$\CC$ relaxes to $\CC'$, on a time scale $\tau_d$, given by (\ref{equ:tau-ell}).  
(For small $c$, the probability that $\CC$ relaxes to
some other local structure is vanishingly small, due to the separation of time scales in the problem.)
After the time lag $\tau_d$, the two configurations behave the same, so all contributions
to $R_\CC-R_{\CC'}$ come from times smaller than $\tau_d$.  The dominant contribution to $R_\CC - R_{\CC'}$ comes
from the spin marked $\times$ in Fig.~\ref{fig:d-sketch}.  
This spin is facilitated throughout the time $\tau_d$
so its contribution to $R_\CC$ is approximately $2c \tau_d$. (This may be shown using an analysis similar to that
leading to (\ref{equ:nr}) above, or the quasi-equilibrium argument
explained at the end of Sec.~\ref{sec:qe-simple}.)  Thus, the enhancement of $\CC$ with respect to $\CC'$ is determined by
\begin{equation}
\Delta V_{\CC'} - \Delta V_{\CC} = 2\nu (R_\CC - R_{\CC'}) + O(\nu^2) \approx 4\nu c\tau_d + O(\nu^2),
\label{equ:pc-prop}
\end{equation}
where the approximate equality 
holds for small $c$ and $2\leq d\lesssim1/c$. (For $d=1$, a similar argument shows that the propensity difference is of order unity: here the dominant contribution comes from the spin at distance $d=1$ itself, which has a flip rate of $1-c\approx 1$, rather than its right neighbour.)
The key point here is that 
the difference in effective potential in (\ref{equ:pc-prop}) depends very strongly on the position of the extra up spin in $\CC'$.
If $d=1$ and the extra up spin is adjacent to an existing one, the difference in effective potential between $\CC$ and $\CC'$
is $\nu\cdot O(1)$, which is small on the natural scale of $\nu$. 
But if (for example) $d=1+2^b$ then the enhancement diverges as $\nu\cdot O(c^{-b})$: configurations where the extra
up spin is far from any existing spin are very strongly enhanced if $b>1$. 

This strong dependence of $\Delta V_\CC$ on the relative positions of the up spins in $\CC$ is
the mechanism for the repulsive correlations in $C(x)$ shown in Fig.~\ref{fig:gr-pd}.  Configurations where the spins are
spaced out have large propensities for activity $R$, since the up spins increase activity, and widely-spaced up spins
persist for longer time within the system.  Hence, the 
effect of the bias $\nu$ is to favour configurations $\CC$ with long-lived active regions, since these give the largest
contributions to $R_\CC$.

\subsection{Response of $p(d)$ to the bias $\nu$}
\label{sec:pd_enh}

To further elucidate this effect, we show how $\nu$ affects the domain structure in the system, by calculating
the response of $p(d)$ at leading order in $\nu$.
Recall that $p(d)$ can be written as in Eq.~(\ref{equ:def-pd}) above,
in terms of the observable $D_{i,i+r}$ defined in
(\ref{equ:UD}). The linear response relation~(\ref{equ:LR}) then gives 
for the enhancement of $p(d)$ at first order in $\nu$
\begin{eqnarray}
  \frac{p(d)}{p^0(d)}
  &= 1 + 2\nu 
\left\langle \left( \frac{n_i  D_{i+1,i+d-1} n_{i+d}}{\langle n_i D_{i+1,i+d-1} n_{i+d}\rangle_0} - \frac{n_i}{\langle n_i \rangle_0} \right) \delta R \right\rangle_0   + O(\nu^2) 
\label{equ:pdpert1}
\end{eqnarray}
The right hand side of (\ref{equ:pdpert1}) is straightforward to
evaluate numerically: 
see Fig.~\ref{fig:gr-pd}(c).

For small $c$, the right hand side of
(\ref{equ:pdpert1}) may be estimated following the discussion of Sec.~\ref{sec:pert-rho}.
The first term in the average in (\ref{equ:pdpert1}) has contributions of the form
\begin{equation}
\frac{1}{\langle n_i D_{i+1,i+d-1} n_{i+d}\rangle_0}
\int \mathrm{d}t\, \langle  [ n_i(t') D_{i+1,i+d-1}(t') n_{i+d}(t') ] \delta r_j(t) \rangle_0,
\label{equ:nifr}
\end{equation}
for which the largest contributions come from $j=i+1$ and $j=i+d+1$:
these sites are adjacent to the spins that are up at time $t'=\tobs/2$ 
and hence facilitated.
If $j=i+1$, Eq.~(\ref{equ:nifr}) gives $\approx 2c\tau_0$ as in Sec.~\ref{sec:pert-rho}; the
amplitude $2c$ comes from the quasiequilibration rule discused in Sec.~\ref{sec:qe-simple}. However, there is an analogous
contribution from the $\langle n_i \delta R\rangle_0$ term in (\ref{equ:pdpert1}) which exactly cancels this effect.
Evaluating (\ref{equ:nifr}) when $j=i+d+1$, one obtains $\approx 2c\tau_d$, which is an alternative derivation of
the enhancement considered in Sec.~\ref{sec:rel_enh}.
In addition, there are contributions to (\ref{equ:pdpert1}) of the form of (\ref{equ:nifr}), with sites 
$j$ between $i$ and $i+d$.  These spins are down and unfacilitated
at time $t=0$, and the typical time scale for spin $i+a$ to become facilitated is $\tau_a$.  
Such a spin therefore contributes $-\tau_a\langle r \rangle = -2c^2\tau_a$ to (\ref{equ:pdpert1}).
This contribution is smaller than the contribution of site $i+d+1$, but if $d$ is large then there are many down spins between $i$ and $i+d$, 
and these contributions become significant.
If we use the coarse (over-)estimate $\tau_a \approx \tau_d$, the total
contribution from down-spins is $-2(d-1)c^2\tau_d \approx
(-cd)(2c\tau_d)$. Since $\tau_d\approx (cd)\tau_0$, this negative
contribution scales with $d^2$ and thus dominates for large $d$ over the positive
contribution from the up-spin at $j=i+d+1$, $2c\tau_d$, which scales
linearly with $d$. A more refined estimate accounting for the
$a$-dependence of the contributions from the internal down-spins at
$i+a$ only changes the prefactor of the leading $d^2$-dependence of
the negative linear response contribution.

Combining all these results, and taking $c\to0$, we expect
\begin{equation}
  \frac{p(d)}{p^0(d)} \simeq 1 + A_d \nu/c^{\alpha_d-1}+ O(\nu^2), \qquad 2\leq d \lesssim 1/c;
\label{equ:pd-pert-result1}
\end{equation}
For domains of size $1$, one has  (by arguments analogous to those in Sec.~\ref{sec:rel_enh})
\begin{equation}
 \frac{p(d=1)}{p^0(d=1)} \simeq 1+A_1\nu + O(\nu^2), 
 \label{equ:pd-pert-result-qe}
\end{equation}
and for large domains
\begin{eqnarray}
\frac{p(d)}{p^0(d)} &\simeq 1+4\nu (1-cd)c\tau_d + O(\nu^2)
\nonumber \\
                  & \simeq 1-A_d\nu\tau_0c^2d(cd-1)+ O(\nu^2), \quad d \gtrsim 1/c.
\label{equ:pd-pert-result2}
\end{eqnarray}
Here, all the $A_d=O(1)$ as $c\to0$. 

We observe that for large domain sizes $d \gg 1/c$, the linear response result diverges with $d$, so the result
is necessarily applicable only in a range of $\nu$ that becomes vanishingly small as $d\to\infty$.
However, such large domains are extremely rare in any case and the effect of $\nu>0$ is to \emph{suppress} 
them very strongly (recall Fig.~\ref{fig:gr-pd}).
In this case,
the breakdown of perturbation theory in the calculation of $p(d)$ does not lead
to any apparent non-perturbative effect in observables like $r(\nu)$ or $C(x)$.  
On the other hand, if we bias to lower than average activity by taking $\nu<0$, 
large domains are \emph{enhanced} non-perturbatively as $d\to\infty$: this effect drives
the phase transition into the inactive state~\cite{kcm-transition}, with large domains predominating for all $\nu<0$.

\section{Hierarchy of responses at small $c$, and link to aging}
\label{sec:hier}

\begin{figure}
\hspace{2cm} \includegraphics[width=11cm]{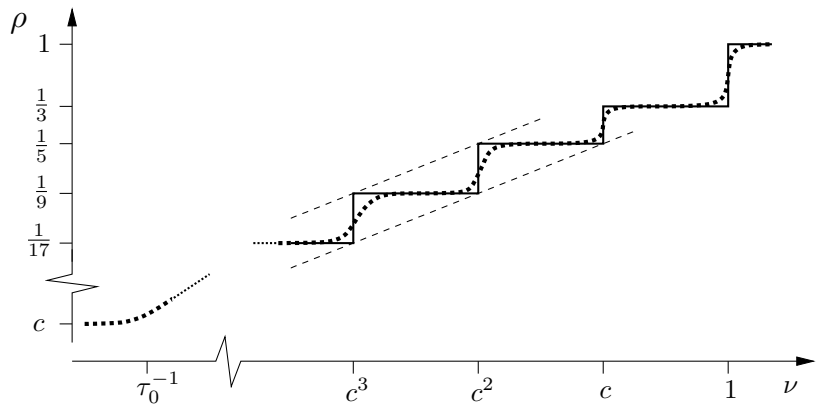}
\caption{
Sketch showing how the density of up spins $\rho = \langle n_i \rangle_\nu$ depends on $\nu$, for very small $c$. Both axes are logarithmic. The solid line shows the limiting behaviour as $c\to0$, taken at fixed $(\ln\nu)/(\ln c)$, while the dashed line shows the expected behaviour for a system with small positive $c$. The effect of the bias $\nu$ becomes significant when $\nu \approx \tau_0^{-1}$, the inverse bulk relaxation time. As $\nu$ is increased further towards unity, the system eventually responds in a sequence of steps. We expect a weak dependence on $\nu$ whenever $c^{b} \ll \nu \ll c^{b-1}$ for integer $b$, leading to plateaus in $\rho$. Within each plateau ($b\geq1$), our conjecture is that the spacing between up spins converges to $1+2^b$ as $c\to0$, so $\rho\to\frac{1}{1+2^b}$. For large $b$, the plateaux are then bounded by $\rho=2^{-b-1}$ and $\rho=2^b$, corresponding to power-law behaviour $\rho\sim 2^{-(\ln\nu)/(\ln c)} \sim \nu^{T\ln 2}$ (dashed lines). From the quasiequilibrium argument, one expects also $r(\nu) \approx 2c \langle n_i \rangle_\nu = 2c\rho(\nu)$: when comparing this prediction with Fig.~\ref{fig:rnu}, we emphasise that those numerical results are still rather far from the small-$c$ limit.
%
%
} \label{fig:plat}
\end{figure}


The results of Section~\ref{sec:pert} can be summarised by 
(\ref{equ:pd-pert-result1}-\ref{equ:pd-pert-result2}), which encode a hierarchy of responses to the bias $\nu$.
When $c$ is small, the linear responses for different $d$ scale as $\nu/c^{\alpha_d-1}$.  Estimating the range
over which this linear response prediction is valid is not a trivial task, because the responses
are divergent for large $d$, consistent with the phase transition to an inactive state as $\nu$ becomes
negative.  However, based on the numerical results of Sec.~\ref{sec:num} and the linear response calculation, 
we can formulate a picture that captures the main features of the responses to the bias, both linear and nonlinear.

The main idea is a generalisation of the quasiequilibrium relation discussed in Section~\ref{sec:qe-simple}.
That argument indicates that $p(d=1)\approx c$ as long as $\nu\ll1$, regardless of whether $\nu$ is small compared
with other powers of $c$.  Put another way, nonlinear responses in (\ref{equ:pd-pert-result-qe}) do not set in until $\nu\simeq 1$.
We are proposing here that similar quasiequilibrium conditions hold for larger $d$, and small $c$.  In particular,
for $\nu/c^{b-1} \ll 1$, domains of sizes $1\leq d \leq2^b$ are weakly affected by the bias, so that $p(d)=O(c)$
as $c\to0$.  This corresponds to the assumption that the linear response prediction
(\ref{equ:pd-pert-result1}) remains accurate until the relative linear response correction becomes of order unity. [More precisely, we expect that  
nonlinear corrections to $p(d)$ are at most of absolute size $O(c)$ in this regime, not $O(1)$.]
On the other hand, the perturbative expansion indicates that larger domains ($d>2^b$) 
have much larger responses which must be nonlinear in nature: we expect 
that $p(d)$ has a non-zero limit for those domain sizes, 
so that the mean domain size (and hence $r(\nu)$ and $\langle \rho \rangle_\nu$)
remain finite as $c\to0$. 

To illustrate this effect, recall Fig.~\ref{fig:sc2} which shows data for $\nu=c^2$, as $c$ is decreased. One sees that
 $p(d)$ for $d\leq 4$ decreases as $c$ is reduced, consistent with the expectation that it tends to
 zero as $c\to0$; on the other hand $p(d)$ has a non-zero limit for $d\geq 5$.  
%
%
We identify an emergent length scale $d^*=5$: domains smaller than $d^*$ have a vanishing probability in the relevant
limit, domains of size close to $d^*$ are very likely, while larger domains have finite (typically smaller) probabilities.
In general, if  $\nu=c^b$ for some integer $b$, and we consider the limit $c\to0$, then one expects a similar situation with
a length scale $d^*=1+2^b$.
 
It is also useful to consider the case $c^b \ll \nu \ll c^{b-1}$ with $c\to0$.  
Then, one expects small domains ($d\leq 2^b$) to be quasi-equilibrated
with $p(d)\simeq c$, while larger domains ($d>2^b$) should have $p(d)=O(1)$, weakly dependent on $\nu$.  In fact, our numerical results
indicate that almost all domains will be of size $d=1+2^b$ in this limit, leading to a finite density of up spins $\langle n_i\rangle_\nu = 1/(1+2^b)$.
This allows the system to maximise the escape rate $r$, subject to the quasiequilibrium constraint on smaller domains.
The simplest case is the limit $c \ll \nu \ll 1$ where this analysis predicts $\langle n_i\rangle_\nu \approx \frac13$.  This behaviour is consistent
with Fig.~\ref{fig:rnu}(c), where $\rho(\nu) = \langle n_i\rangle_\nu$ increases quickly to a value close to $\frac13$ at $\nu\approx c =0.1$
before increasing more weakly for large $\nu$.  
The value of $c$ is not small enough to saturate the limit and establish a clear
plateau in $r(\nu)$ but the numerical results are certainly consistent with the picture proposed here. 
Returning to the general argument, Fig.~\ref{fig:plat} summarizes the predicted hierarchy of responses in a sketch. 
The plateau structure in $\rho(\nu)$ and its power-law asymptote $\rho\sim \nu^{T\ln2}$ are intriguingly similar to the one observed in the out-of-equilibrium aging dynamics of the East model~\cite{Sollich-Evans}, with $\nu$ playing the role of the inverse age.
A similar generalisation of quasiequilibrium to that described here 
can also be observed in equilibrium dynamics, through an analysis of 
metastable states~\cite{jack-patch13}.

\section{Variational approaches}
\label{sec:var}

\newcommand{\ut}{\tilde{u}}
\newcommand{\dVt}{\Delta\tilde{V}}
\newcommand{\Fvar}{F^\mathrm{var}}
\newcommand{\pvar}{p^\mathrm{var}}
\newcommand{\Vvar}{\Delta V^\mathrm{var}}

To investigate the response of the system to $\nu$ beyond first order, we exploit a variational method,
which relies on the time-reversal symmetry of the $\nu$-ensemble.
The master operators $\WW_K(s)$ and $\WW_R(\nu)$ may be symmetrised, as described in Sec.~\ref{sec:large-dev}.
Hence (see~\ref{app:block} and Refs.~\cite{kcm-transition,Jack-Sollich-PTP}),
one may obtain the effective potential $\Delta V_\CC$ by minimising the variational `free energy' (per site):
\begin{equation}
  F(\dVt) =  -N^{-1} \frac{ \sum_{\CC,\CC'} \ee^{-\dVt_{\CC'}/2} [\WW_R(\nu)]_{\CC',\CC}  \ee^{-\dVt_{\CC}/2} p^0_{\CC}}
                          {  \sum_{\CC} \ee^{-\dVt_{\CC}} p^0_\CC }  ,
  \label{equ:Fvar}
\end{equation}
where $\dVt$ is a variational estimate of the effective potential,
$[\WW_R(\nu)]_{\CC',\CC}$ is a matrix element of the operator $\WW_R(\nu)$, and $p^0_\CC=\ee^{-\beta\sum_i n_i}/(1+\ee^{-\beta})^N$ 
is the equilibrium probability of configuration $\CC$.  On minimising $F(\dVt)$ over all the $\dVt_\CC$, the minimal value of $F$
is equal to the dynamical free energy $\psi_R(\nu)$, and $\Delta \tilde V_\CC$ is equal to the effective potential $\Delta V_\CC$.  Hence,
if a suitable exact parameterisation of $\Delta \tilde V_\CC$ may be found, one may obtain the effective potential by minimising
$F(\dVt)$.  More typically, one makes an approximate parameterisation of the effective potential, and minimises
$F$ with respect to the variational parameters.  
For a given parameterisation of $\Delta \tilde V_\CC$ (``trial potential''), we denote the minimal value of $F$ by $\Fvar$
and the corresponding estimate of $\Delta \tilde V_\CC$ by $\Delta V^\mathrm{var}_\CC$.

We note in passing that an alternative to this variational approach would be to use a density-matrix renormalisation
group method (see e.g.~\cite{gorissen09}), which is related to a variational search over matrix product states~\cite{schollwoeck11}.
However, the advantage of (\ref{equ:Fvar}) is that it has a clear interpretation as a variational search
in a given space of effective potentials $\Delta V$.

\subsection{Trial potential functions}

We first describe three trial potentials that we have investigated.  

\subsubsection{Block model.}
\label{sec:block}

A completely general trial potential should include all possible $m$-body interactions of all ranges, 
To approximate this, we consider interactions within `blocks' of size $B$.  
The potential includes $m$-body interactions up to $m=B$, with a maximal interaction range of $B-1$.
It also permits a transfer-matrix representation of the probability distribution over configurations $\CC$ in the $\nu$-ensemble.  
The idea is to consider a block of spins, ${\cal B}_{i}=(n_i,n_{i+1},\dots,n_{i+B-1})$, 
and that each possible block configuration has its own contribution to the trial potential.  That is, if $e_{\cal B}({\cal B}_i)$
is an indicator function, equal to unity when block ${\cal B}_i$ has the specific configuration ${\cal B}$ and zero otherwise, then 
$\dVt_\CC = \sum_i \sum_{\cal B} z_{\cal B} e_{\cal B}({\cal B}_i)$, where the $z_{\cal B}$ are variational parameters that determine
the block probabilities.  

There are $2^B$ trial weights $z_{\cal B}$, but these in fact provide an overcomplete basis for the possible interactions in $\dVt$,
because the numbers of blocks in the $2^B$ different states are not independent.  For example, if $N_{010}=\sum_i \nbar_i n_{i+1} \nbar_{i+2}$
is the number of blocks with configuration `$010$' (and similarly for other block configurations) then one may use $\nbar=1-n$ to write
$N_{010}=\sum_i[ n_{i+1}\nbar_{i+2} - n_{i} n_{i+1}\nbar_{i+2} ] = \sum_i[ n_{i+1}\nbar_{i+2} n_{i+3} + n_{i+1}\nbar_{i+2} \nbar_{i+3} - n_{i} n_{i+1}\nbar_{i+2} ]  = N_{101} + N_{100} - N_{110}$. Here the last equality involved a relabelling within the summation, which relies on the periodic boundaries
of the system.  In this way, the numbers of all blocks of length $B$ that begin with a down spin (`0') can be expressed exactly in terms of numbers of blocks that begin 
with an up spin (`1'). There are $2^{B-1}$ such numbers, and it is then not difficult to see that one can equivalently specify the numbers of all blocks of length $1$ to $B$ that start {\em and} end with a 1. (E.g.\ for $B=2$ one can use $N_{1}=N_{11}+N_{10}$ and $N_{11}$ instead of $N_{10}$ and $N_{11}$.)
Using this latter representation, a general trial potential for the block model with block length $B=2$ can be written as
\begin{equation}
\dVt_\CC = \sum_{i} h n_i  + J n_i n_{i+1}    ,
\label{equ:ut-block}
\end{equation}
which includes a field $h$ and a two-body Ising-like coupling as variational parameters.  
For higher $B$ one has in addition all interaction terms up to range $B-1$ and involving up to $B$ spins.

In~\ref{app:block}, we show how the variational free energy in (\ref{equ:Fvar}) may be calculated for this model, given the weights
$z_{\cal B}$.  The method relies on a transfer matrix representation of $\ee^{-\dVt_\CC}$. This allows $F$ to be minimised (numerically) 
over the $z_{\cal B}$, leading to an estimate $\dVt^\mathrm{var}$ for the effective interactions.

\subsubsection{$p_d$ model.}
\label{sec:var-pd}

The block model is a general variational ansatz, but the number of variational parameters increases exponentially with the 
maximal interaction range $B-1$.  In the following, this will limit our numerical results to $B\leq 6$.  
As we have already discussed, Fig.~\ref{fig:gr-pd} indicates that the true effective potential $\Delta V_\CC$
includes rather long-ranged interactions, which would require much larger values of $B$ to capture them.

For this reason, we have designed 
trial potentials in order to account for the particular structures that occur in the East model.
Instead of including all interactions up to range $B$,  these potentials 
include a specific subset of long-ranged interactions. 
In particular, we concentrate on the `domains' discussed above and
construct a $\Delta \tilde V_\CC$ that depends only on
the sizes of these domains.
Thus, this trial potential will be accurate if all information about structure in the $\nu$-ensemble is contained
in the distribution $p(d)$ shown in Fig.~\ref{fig:gr-pd}.  Formally, we include specific $m$-body interactions of all ranges
\begin{equation}
\dVt_\CC = \sum_{i,d} z_d n_i D_{i+1,i+d-1}  n_{i+d} ,
\label{equ:ut-pd}
\end{equation}
where the $z_d$ are variational parameters associated with the possible domain sizes.  Given these weight factors, 
one may derive the distribution of domain sizes $p(d)$ associated with this variational ansatz.  In fact, it is convenient
to work directly with this distribution, which we denote by $p_d$ to distinguish its status as a variational parameter from a measured $p(d)$.

The variational free energy is
\begin{equation}
F_d =\frac{1}{\sum_d d p_d} \Big\{  (1-\nu)[c + (1-2c)p_1] - 2\sqrt{c(1-c)} \sum_{d\geq 2} \sqrt{p_1 p_{d-1} p_d}  \Big\} .
\label{equ:Fd}
\end{equation}
The derivation of this free energy is discussed in~\ref{app:pd}: it is to be minimised subject to the normalisation
constraint $\sum p_d = 1$.  

In principle the sums over $d$ in (\ref{equ:Fd}) run over all $d\geq1$ and $d\geq 2$ respectively, but for numerical work
we truncate the sums at a cutoff $d^*$ by assuming $p_d=0$ for $d>d^*$: 
domains much larger than $1/c$ are extremely rare in the system so the results depend negligibly on $d^*$. 

\subsubsection{$p_{fe}$ model.}
\label{sec:var-fe}

\begin{figure}
\hspace{2.5cm}\includegraphics[width=7cm]{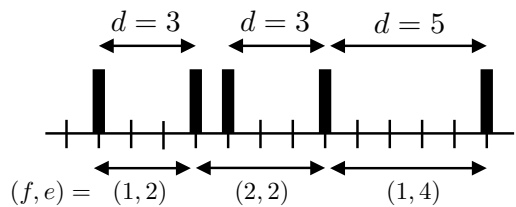}
\caption{Sketch showing how the domains are identified in the $p_{fe}$ and $p_d$ models.  The (unlabelled)
domain of size $d=1$ in the $p_d$ representation is combined with the adjacent domain
to form an composite domain with $(f,e)=(2,2)$ in the $p_{fe}$ model.
} \label{fig:pfe-sketch}
\end{figure}

The $p_d$ model gives useful results, but we will find that it does not account accurately for the quasiequilibrium conditions 
described in Sec.~\ref{sec:hier}.  This shortcoming limits its accuracy, so we 
discuss one systematic improvement to the $p_d$ model, which captures
some features of this quasiequilibration.

Recall that in the $p_d$ model, the system is divided into domains that start at each up spin, and each domain is assumed to be
independent.  Each domain then consists of a single up spin, followed by a block of $d-1$ down spins.
An alternative and more general assumption is to start domains at each occurrence of the block `$01$'.  The situation is illustated
in Fig.~\ref{fig:pfe-sketch}. Each domain
consists of a block of up spins, followed by a block of down spins.  
The domain state is specified by $2$ numbers $f,e$, so that there are $f$ (``full'') up spins and $e$ (``empty'') down spins.
Fig.~\ref{fig:pfe-sketch} shows that the domains in this `$fe$'-representation
are in general larger than those in the $p_d$-model representation.  We then construct an effective potential on the assumption 
that the $fe$-domains are independent:
\begin{equation}
\dVt_\CC = \sum_{i,f,e} z_{fe} \nbar_{i-1} \cdot U_{i,i+f-1} \cdot D_{i+f,i+f+e-1} \cdot n_{i+f+e} ,
\end{equation}
where the $z_{fe}$ are the variational parameters.  We refer to this model as the `$p_{fe}$ model'.
As with the $p_d$ model, it is more convenient to work with the probability distribution over the domains, which we 
denote by $p_{fe}$.  One can show that the $p_d$ model corresponds to the special case $p_{fe} = p_1^{f-1} p_{e+1}$,
which emphasises that the $p_{fe}$ model is a generalisation of the $p_d$ model.  In particular, the $p_{fe}$ model
allows the length $e$ of a down-block
to depend on the number of adjacent up spins to its left.

The derivation of the variational free energy for the $p_{fe}$ model is discussed in~\ref{app:pfe}: the result is
\begin{eqnarray}
F_{fe} = & 
\frac{1}{\sum_{fe}(f+e)p_{fe}} \cdot \Big\{  (1-\nu)\sum_{fe} p_{fe} [c + (f-1)(1-c)] 
\nonumber \\
      & - 2\sqrt{c(1-c)} \Big[ \sum_{fe} \sqrt{p_{f,e+1}p_{f+1,e}}  
      + \sum_{f,f',e} \sqrt{p_{f'1}p_{fe}p_{f+f'+1,e}} \Big] \Big\}  ,
\label{equ:Ffe}
\end{eqnarray}
The minimisation is subject to the normalisation constraint
$\sum_{f,e}p_{fe}=1$. Numerically we again model $p_{fe}$ explicitly
only up to cutoffs $f^*$ and $e^*$. These cutoffs cannot be made too
large because the number of variational parameters is now $f^* e^*$.
To soften the impact of the cutoffs we therefore do not set $p_{fe}$
directly to zero beyond the cutoffs, but assume an exponential tail
instead that is obtained by linear extrapolation of $\ln p_{fe}$.

\subsection{Variational results}

\begin{figure}
\hfill \includegraphics[width=7cm]{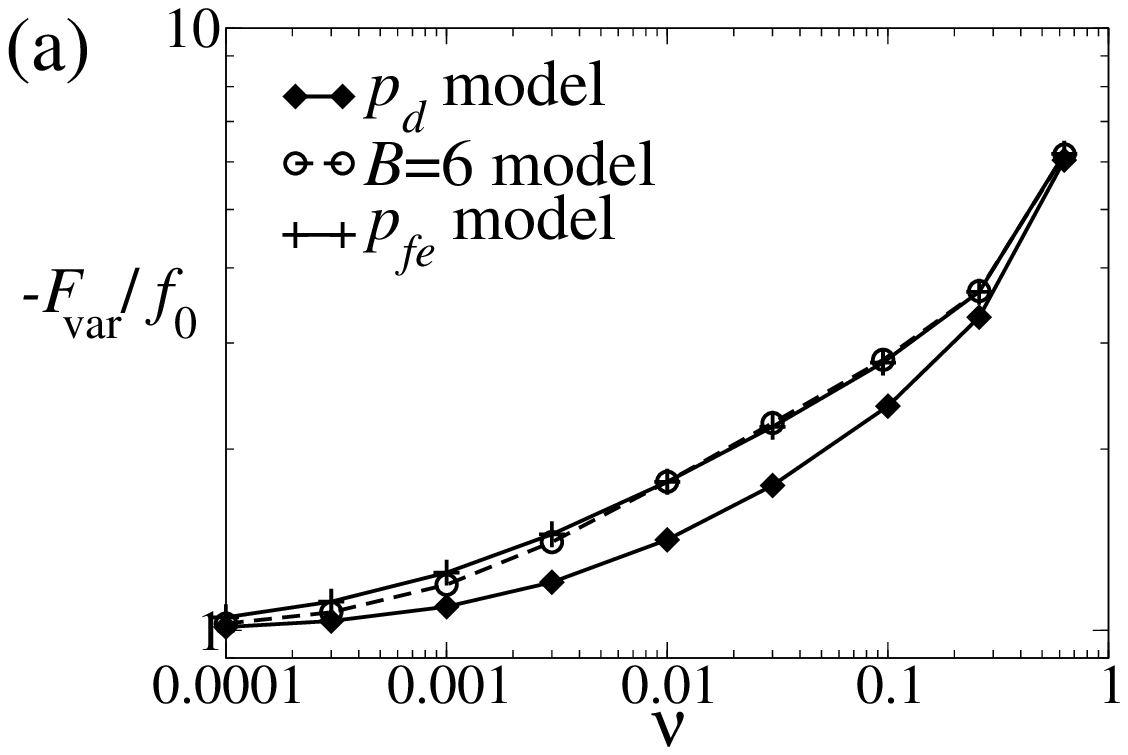}
\includegraphics[width=7cm]{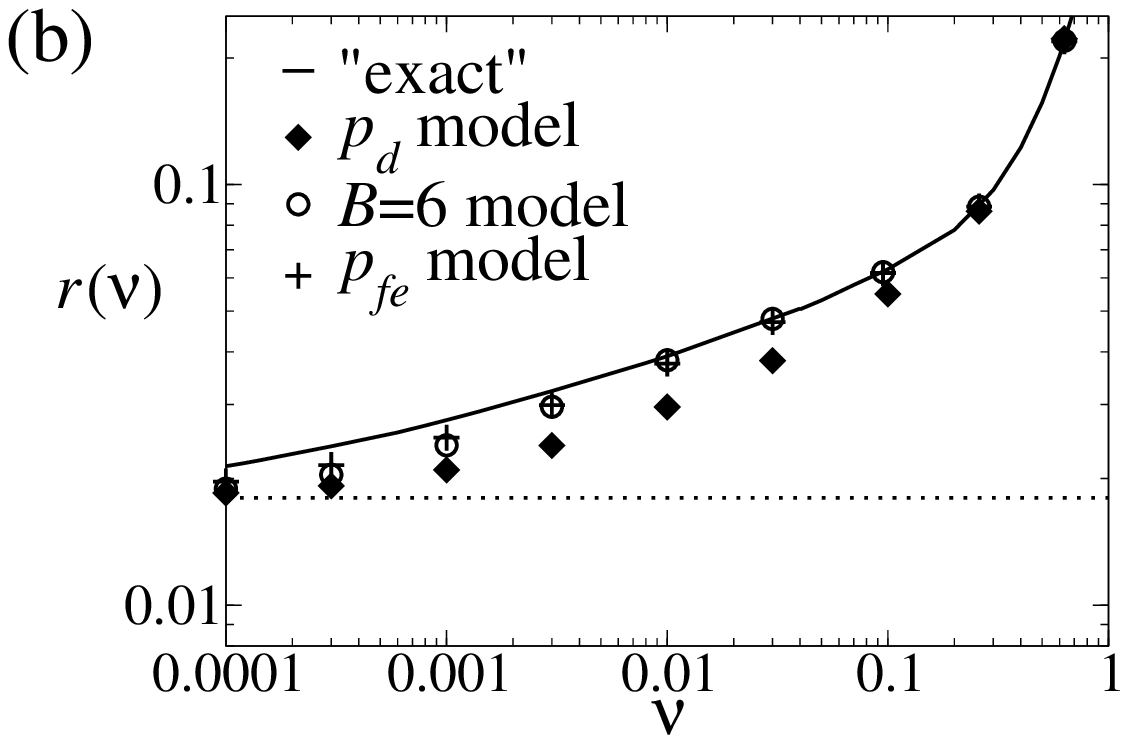}
\caption{
  (a) Variational free energy, $\Fvar$ at $c=0.1$, using different trial potentials.  
  Since $\Fvar$ varies over several orders of magnitude, we plot
  $-\Fvar / f_0$, where $f_0= \nu r(0)$. One has $-\Fvar / f_0\to1$ as $\nu\to 0$, so that $\Fvar = -f_0 + O(\nu^2)$ at small bias.
  A larger number in this representation indicates a better trial potential:
  the $p_{fe}$ model and the $B=6$ block model perform significantly better than the (simpler) $p_d$ model.  
  (b)~Variational estimates of $r(\nu)$ compared with the numerical results shown in Fig.~\ref{fig:rnu} (labelled ``exact'').
  The dotted (horizontal) line indicates $r(0)=2c^2(1-c)$.
}
\label{fig:rnuvar}
\end{figure}

We have used numerical minimisation to obtain results for the three trial potentials, at $c=0.1$.
 Fig.~\ref{fig:rnuvar}  shows the values of $F^\mathrm{var}$ that we obtained, and the corresponding 
estimates for $r(\nu)$.   These are compared with the numerical results shown in Fig.~\ref{fig:rnu}. 
In general, the $p_{fe}$ model and the $B=6$ block model seem to capture the data
quite well, while the $p_d$ model gives less good agreement.  It is also clear from $r(\nu)$ that
the variational models perform best for larger $\nu$, with significant deviations for smaller $\nu$.

\begin{figure*}
\hfill 
\includegraphics[width=7cm]{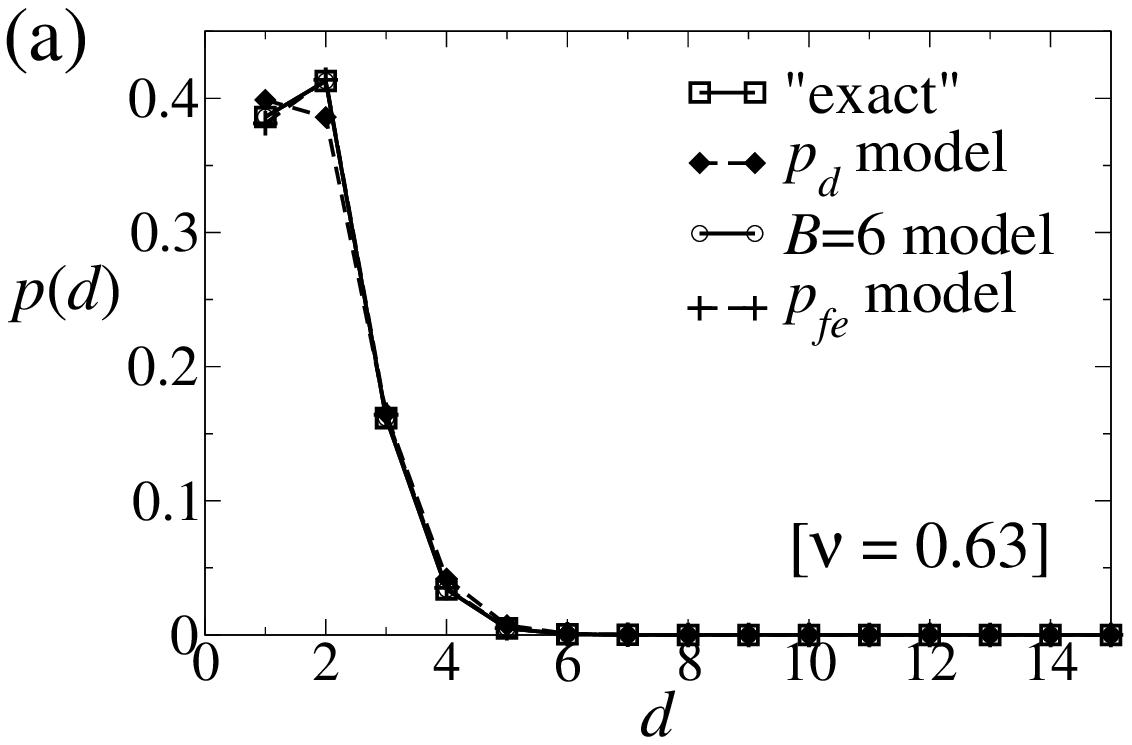}
\includegraphics[width=7cm]{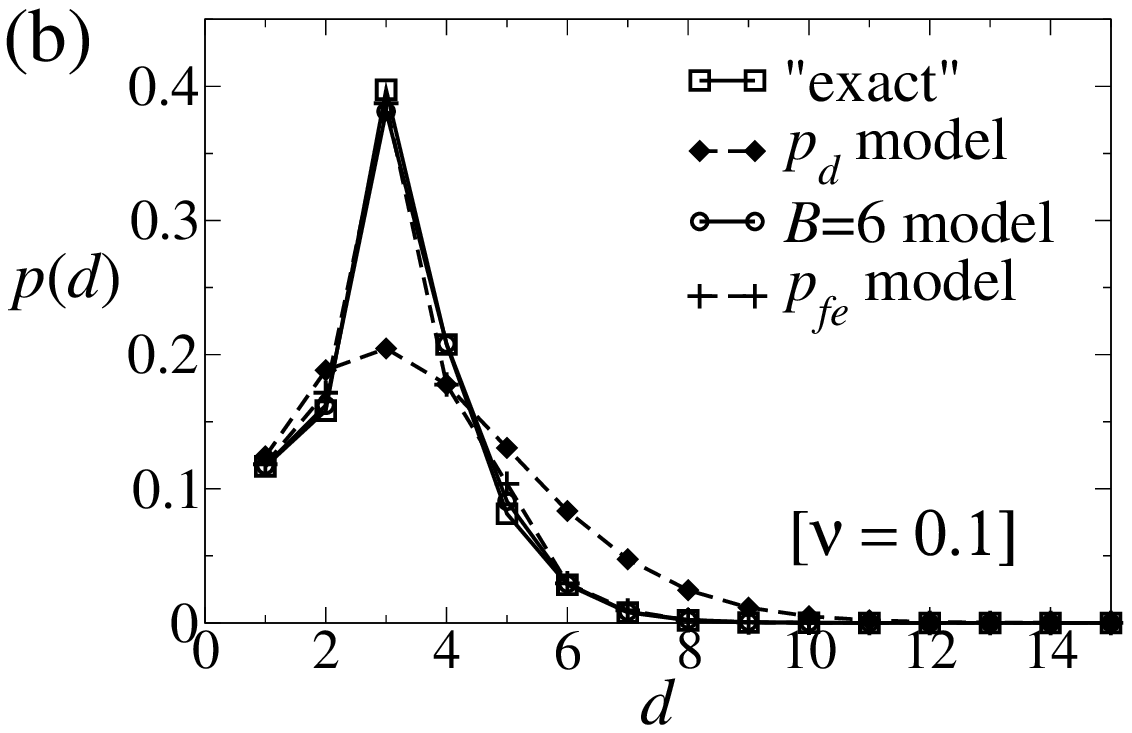}\par
\hfill 
\includegraphics[width=7cm]{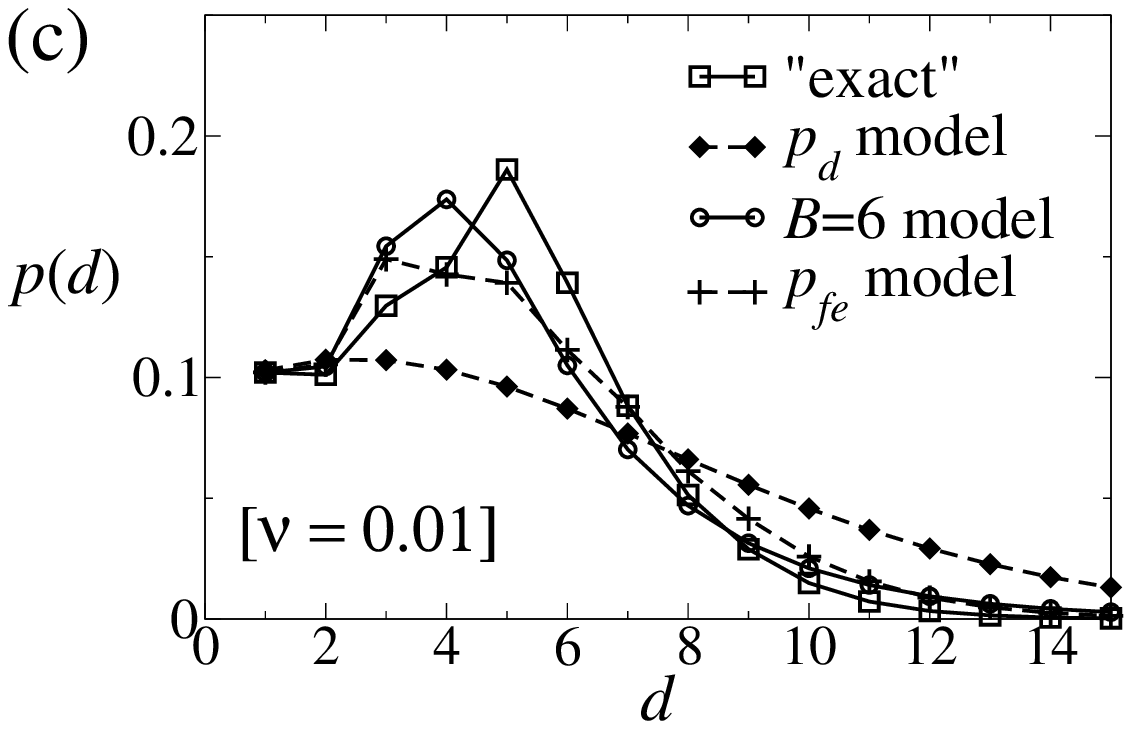}
\includegraphics[width=7cm]{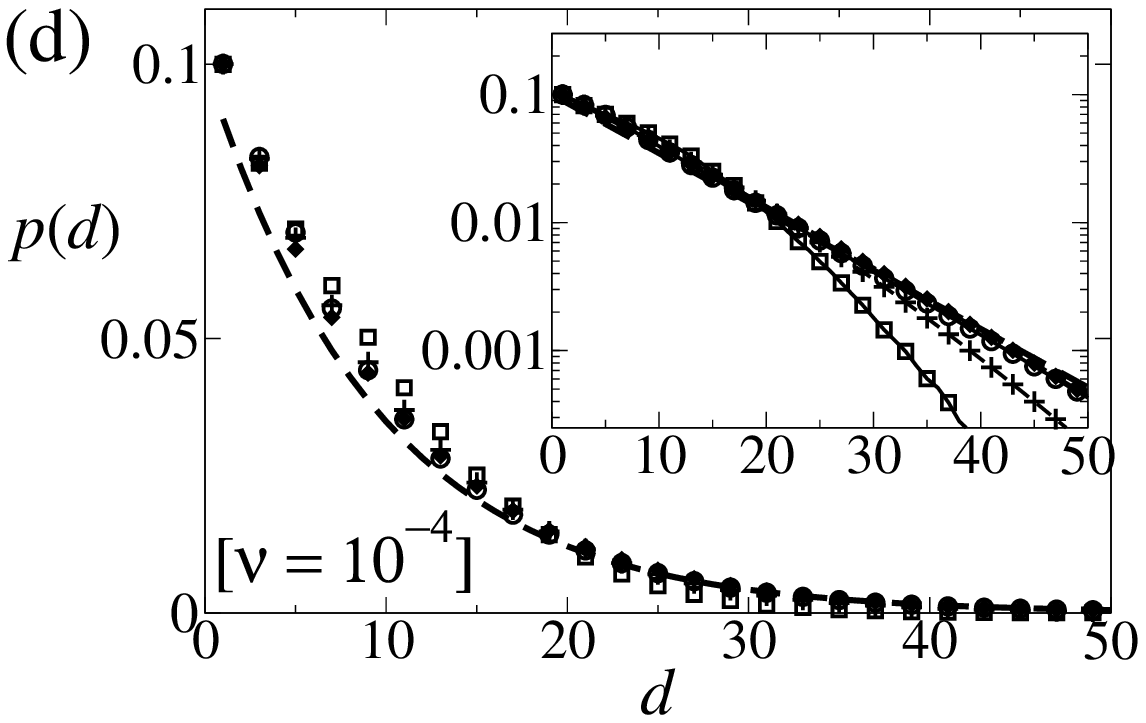}
\caption{(a-d) Comparison between the domain-size distributions $p(d)$ obtained numerically by TPS or exact diagonalisation 
(labelled ``exact'')
with those obtained from (approximate) variational analyses.  
All results are at $c=0.1$, and for four different values of $\nu$ as shown in each plot.  
The variational
approaches are most effective for larger $\nu$ when domains are typically short. 
%
%
In panel (d),
the unbiased distribution $p^0(d)$ is shown as a dashed line.  The inset
shows the same results on a logarithmic scale, emphasising the differences in the tails of the distribution.
For this value of $\nu$, the $p_d$ model and the $B=6$ model
give similar results, and are both close to the unbiased distribution $p^0(d)$.  (The legend is omitted:
symbols are the same as in all other panels.)}
\label{fig:pdvar}
\end{figure*}

Since the minimisation yields the effective interaction potential $\Vvar_\CC$, we are also able to calculate
variational predictions for one-time quantities in the $\nu$-ensemble.  As a 
stringent test of these variational distributions,  
Fig.~\ref{fig:pdvar} shows the estimates for $p(d)$ that we obtain from the variational treatment, 
compared with numerical results from Sec.~\ref{sec:num}.  For the largest $\nu$ ($=0.63$), all three models
describe the data quite accurately, although deviations are apparent for the $p_d$ model.
Nearly all domains in this system are short, so one can expect a relatively simple effective
interaction to accurately describe these data.

For $\nu=0.1$, there is clear structure in the system, including a most probable domain size of $d=3$,
 which is captured quite accurately by
both the $p_{fe}$ model and the $B=6$ block model, although not by the $p_d$ model.  
We recall from Sec.~\ref{sec:pert} that for $\nu\approx c$,
one expects domains of lengths $d\leq 2$ to have probabilities of order $c$, while
larger domains have probabilities of order unity.  This is consistent with the most likely domain size of $d=3$.

For smaller $\nu$, the structure in $p(d)$ becomes more complex, and even the $p_{fe}$ and $B=6$ models
fail to accurately describe the effective interactions in the system.  We note in particular that
$\nu=0.01$ corresponds to $\nu=c^2$ for this case, in which case Sec.~\ref{sec:pert} predicts that $p(d)$ should be of
order unity for $d\geq 5$, but of order $c$ for $d\leq 4$.  It is apparent that the $p_{fe}$ and $B=6$ models
fail to capture this aspect of the linear response to the system.  Finally, we note that for very small $\nu\approx 10^{-4}$,
these variational models significantly underestimate the suppression of large domains.  For the block model,
it is easily shown (see~\ref{app:pd-exp}) that $p(d)$ must decay exponentially for $d>B$, in contrast to the faster decrease
found in our numerically exact results.

It is clear from the numerical results in Figs.~\ref{fig:rnuvar} and \ref{fig:pdvar} that the $p_d$ model gives quite a crude description
of the effective interactions in the system.  However, this model can be studied analytically, with several useful results, which we summarize briefly to conclude this section. Looking at the linear response for small $\nu$, one finds for $p_1$ a positive relative correction of $O(\nu)$ in line with the notion of quasi-equilibrium for small $\nu$, while for all other $p_d$ the relative correction is $O(\nu/c)$. With increasing $d$ the correction becomes negative and its amplitude grows, so that the model captures at least qualitatively the large-$d$ divergence of the perturbative correction shown in Fig.~\ref{fig:gr-pd}(c). For nonzero $\nu$ one can show that $p_d$ must decay faster than exponentially, indicating the presence of long-ranged effective interactions that will make block models with fixed block lengths poor approximations. 
Finally one can look at the $c\to 0$ limit of the $p_d$ model at fixed small (but non-zero) $\nu$. One finds $p_1=O(c)$, consistent again with quasi-equilibrium, while $p_d=O(1)$ for $d\geq 2$. The predicted activity is $r(\nu)=c\cdot f_r(\nu)$ with $f_r$ of $O(1)$. Assuming that quasiequilibrium along the lines of (\ref{equ:qe-pred}) holds, one infers that $\langle \rho \rangle_\nu = f_\rho(\nu)$ with $f_\rho(\nu) \approx \frac12 f_r(\nu)$ of $O(1)$: the density of up-spins remains finite, in contrast to the equilibrium case $\nu=0$ where $\langle \rho\rangle_0=c$ vanishes as $c\to 0$. From the discussion of Sec.~\ref{sec:hier}, one would expect that $f_\rho(\nu\to0)=1/3$, so $f_r(\nu\to0) = 2/3$. The $p_d$ model predicts correctly that $f_r$ and $f_\rho$ are of order unity as $c\to0$, but gives a rather poor estimate of the shapes of the function, yielding e.g. $f_r(\nu)\sim \nu^{1/2}$ for small $\nu$.

%
%

\subsection{Limitations of variational schemes: the complex hierarchical response to $\nu$}

The results of Sec.~\ref{sec:pert} suggest a hierarchy of responses to the bias $\nu$, as discussed
in Sec.~\ref{sec:hier}.  The general idea is that the system remains quasiequilibrated on short length scales,
while large length scales respond strongly to the bias.  The linear response of a configuration
to the bias is given by its propensity for activity, $R_\CC$.  We find that $R_\CC$ is dominated by long
time scales in the model, which are typically associated with large scale structures in the configuration $\CC$.

Fig.~\ref{fig:pdvar} shows data at $\nu=10^{-4}$ and $c=0.1$ which indicate that none of the variational models used here
are successful in capturing the long-ranged correlations that appear in this perturbative regime.  While the $B=6$ block model
necessarily excludes long-ranged correlations, the failure of the $p_{fe}$ model indicates that domain sizes alone are not
sufficient to predict the propensities $R_\CC$, so that $\dVt$ necessarily includes interactions between domains of different sizes.

\begin{figure}
\hfill \includegraphics[width=14cm]{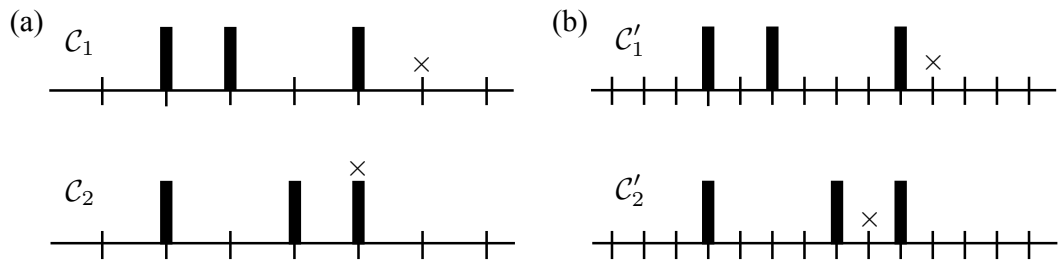}
\caption{
  (a) Configurations $\CC_1$ and $\CC_2$ that illustrate how the propensity depends on correlations
  between domains, and not just on their domain lengths.  (See discussion in main text.) 
   Vertical bars indicate up spins ($n_i=1$) as before.  The $\times$ signs
  mark the sites whose contributions are most relevant when comparing the propensities of these configurations.
  (b) Configurations $\CC_1'$ and $\CC_2'$ indicate that while some relevant correlations between domains
  can be captured within the $p_{fe}$ model, there are still configurations that have different propensities,
  but equal probabilities within the $p_{fe}$ model.
}
\label{fig:var-fail}
\end{figure}

To see how structure among domains is important in determining their propensities (and hence their effective potentials),
it is useful to consider the configurations $\CC_1$ and $\CC_2$
shown in Fig.~\ref{fig:var-fail}(a).
Within the $p_d$ model, these configurations are equally likely.  
However, to estimate their propensities, one follows the analysis of Sec.~\ref{sec:rel_enh} and
first assumes that the least long-lived up spin in each state relaxes quickly to 0.    Then one considers the contributions to $R_{\CC_1}$
and $R_{\CC_2}$ from the spins marked $\times$.  In $\CC_1$, the relevant spin remains facilitated for an $O(c^{-2})$ time scale
while in $\CC_2$ it remains facilitated for an $O(c^{-1})$ time scale.  The contributions to $R_{\CC_1}$
and $R_{\CC_2}$ are therefore $O(1/c)$ and $O(1)$ respectively: the enhancement of $\CC_1$ in the presence of the bias is
much stronger than that of $\CC_2$.  The $p_d$ model cannot capture this difference.  However, 
within the $p_{fe}$ model, configurations $\CC_1$ and $\CC_2$ have independent weights,
proportional to $p_{2,1} p_{1,e}$ and $p_{1,1} p_{2,e}$ respectively.  Thus, the model accounts for their different enhancements 
in the presence of the bias.  The argument that facilitated spins remain
quasiequilibrated implies that $p_{f+1,e} \approx c p_{f,e+1}$, so (for example) $p_{2,1}\approx c p_{1,2}$. 
The numerical results obtained from the $p_{fe}$ model are broadly consistent with this result.  
This effect may be also rationalised in the superspin picture of Ref.~\cite{Sollich-Evans}
if one removes facilitated up spins to leave (relatively) long-lived superspins.  It is the spacing of the superspins that
determines the propensity.  

While the $p_{fe}$ model performs better than the $p_d$ model for configurations $\CC_1$ and $\CC_2$, its main shortcoming may be understood in a similar way.
One generalises the previous argument by multiplying all length scales by two
and all time scales by $1/c$.
The configurations $\CC_1'$ and $\CC_2'$ shown in Fig.~\ref{fig:var-fail}(b) have equal probability within the $p_{fe}$ model but the spin marked with $\times$ in $\CC_1'$ remains facilitated for a time that is $O(c)$ shorter than the marked spin in $\CC_2'$.  Thus, the relative propensities of the configurations
are different, but this effect is missed within the $p_{fe}$ model.  Thus, the $p_{fe}$ model can distinguish configurations that
respond to the bias at $O(\nu)$ from those that respond at $O(\nu/c)$, but it cannot distinguish those that respond at $O(\nu/c^2)$ or greater.

A perfect trial potential would distinguish configurations that respond at $O(\nu/c^b)$, for all the possible values of $b$ discussed in Sec.~\ref{sec:hier}.
However, the point here is that the $p_{fe}$ model specifically allows for $p_d$-domains with $d=1$ to be correlated with other domains to their right.
All other correlations amongst domains are forbidden.
But to resolve the difference in propensity between $\CC_1'$ and $\CC_2'$ in Fig.~\ref{fig:var-fail},
one requires specific correlations between domains with $d=2$ and other domains.  And while including such correlations explicitly
would allow characterisation of configurations which respond at $O(\nu/c^2)$, 
the procedure used to derive $\CC_1'$ and $\CC_2'$
from $\CC_1$ and $\CC_2$ can be repeated to obtain a $\CC_1''$ and $\CC_2''$, one of which will respond at $O(\nu/c^3)$.  
Making this distinction will rely on specific correlations among domains with $d=4$ and greater.  One sees that accounting
specifically for these increasingly complex correlations quickly becomes prohibitive.  The $p_{fe}$-model serves as a useful
indication of the physics at work and the kinds of effective interaction that are expected. We are exploring further improvements to
this trial potential, but these are beyond the scope of this study.

\section{Outlook}
\label{sec:summ}

To end this article,
we discuss which of the features of the analysis here may be generalised to other glassy systems
in the presence of biased activity.  At the perturbative level, we showed that the response of a configuration
depends only on its propensity: this result is general.  The problem of finding the effective interactions for the linear response regime
is therefore equivalent to finding a model that accurately describes the propensity of a configuration.  In the East model,
this requires consideration of structure on quite large length scales, and an accurate description requires identification
of the long-lived superspins in the system, which is a difficult task.  In the general case, we have shown that  variational
calculations can be useful in showing what effective interactions can reproduce the correlations found in biased
states.  However, the trial distributions must be informed by considerable physical insight to yield useful results.

On the other hand, the hierarchy of time scales and the quasi-equilibrium features of the biased East model do simplify
the description of the effective interactions.  If the model is quasiequilibrated on short length scales, this means that effective
interactions on those scales are weak and may be neglected.  Recent work on atomistic model glass-formers~\cite{jack11-stable}
and spin-glass
models~\cite{jack-rom09} does indicate that time-scale separation in biased ensembles can be used to simplify the description of the biased states.
This may well be a useful simplification to guide future studies.

\ack
We thank Fred van Wijland and Juan P. Garrahan for many useful discussions on large deviations,
and emergent structure in ensembles of trajectories.  RLJ was supported by the EPSRC through grant EP/I003797/1.

\begin{appendix}

\section{Calculations using variational trial potentials}

\subsection{Block model}
\label{app:block}

In this section, we describe how the variational free energy in (\ref{equ:Fvar}) is calculated for
the block model of Sec.~\ref{sec:block}.  The block configuration ${\cal B}_i=(n_i,n_{i+1},\dots,n_{i+B-1})$ is a binary string of length $B$,
and a configuration may be specified by its block configurations: $\CC = ( \dots, {\cal B}_i, {\cal B}_{i+1}, \dots )$.  The blocks
are overlapping, so the specification is overcomplete:
the final $B-1$ spins in ${\cal B}_i$ are equal to the first $B-1$ spins in ${\cal B}_{i+1}$, etc.

\newcommand{\ppt}{\tilde p}

We write $\ppt(\CC) = \ee^{-\dVt_\CC-\beta E_0(\CC)}$ where $E_0=\sum_i n_i$ is the energy of the East model, and we
identify $\ppt$ as the (unnormalised) trial probability distribution associated with the trial potential $\dVt$.
From (\ref{equ:ut-block}), one has $\ee^{-\dVt_\CC} = \prod_i \ee^{-z_{{\cal B}_i}}$. It is convenient to write
$\ppt = \prod_i M_{{\cal B}_{i},{\cal B}_{i+1}}$, 
where $M_{{\cal B}_i,{\cal B}_{i+1}}=\ee^{-z_{{\cal B}_i}-\beta n_i}$, recalling that the last $B-1$ spins of ${\cal B}_i$
always coincide with the first $B-1$ spins of ${\cal B}_{i+1}$.  One then generalises $M$ to 
a ``transfer matrix'' of size $2^B \times 2^B$, by setting $M_{{\cal B},{\cal B}'}=0$ if
the last $B-1$ spins of ${\cal B}$ \emph{do not}
 coincide with the first $B-1$ spins of ${\cal B}'$.
With this choice, averages with respect to $\ppt$ can be evaluated as
matrix traces.  E.g., for a periodic chain of length $L$, one has $\sum_{\CC} \ppt(\CC) = {\rm tr}(M^L)$.  We note that
for $B=2$, the block model reduces to the 1$d$ Ising model, which would usually be solved using a $2\times 2$ transfer matrix:
the method presented here uses a $4\times 4$ transfer matrix.  This is less efficient numerically, 
but its generalisation to larger $B$ is simpler.

We now relate the matrix $M$ to the variational free energy $F$ defined in (\ref{equ:Fvar}).  To this end, we symmetrise
the operator in (\ref{equ:Fvar}), noting
that $[\WW_R(\nu)]_{\CC',\CC} p^0(\CC) = \sqrt{p^0(\CC')} [\sum_i \hat{H}_i(\nu)]_{\CC',\CC} \sqrt{p^0(\CC)}$ where 
\begin{equation}
\hat{H}_i(\nu) =  \hat{n}_{i-1} \big[  \sqrt{c(1-c)}(\sig^-_i + \sig^+_i) - (1-\nu)(1-2c)\hat{n}_i - (1-\nu)c   \big]
\end{equation}
is a symmetric (self-adjoint) operator associated with flips of spin $i$.  Since the trial potentials that we consider
are translationally invariant along the chain, (\ref{equ:Fvar}) becomes
\begin{equation}
F = \frac{ - \sum_{\CC,\CC'} \sqrt{\ppt(\CC')} \left[\hat{H}_i(\nu)\right]_{\CC',\CC} \sqrt{\ppt(\CC)} }
             {\sum_{\CC} \ppt(\CC) }
\label{equ:Fvar-H}
\end{equation}
where $[\hat{H}_i(\nu)]_{\CC',\CC}$ indicates a matrix element of $\hat{H}_i(\nu)$.  It is convenient to write
the numerator here as $\sum_{\CC,\CC'}  O_i(\CC,\CC') \ppt(\CC)$, with
\begin{equation}
O_i(\CC',\CC) = \sqrt{\frac{\ppt(\CC')}{\ppt(\CC)}} \left[\hat{H}_i(\nu)\right]_{\CC',\CC} .
\label{equ:Oi}
\end{equation}
We note (i)~that $O_i(\CC',\CC)=0$ unless $\CC$ and $\CC'$ coincide for all spins $j$ except $j=i$,
and (ii)~that $O_i(\CC',\CC)$ depends only on spins $n_{i-B+1},\dots,n_{i+B-1}$ (as long as $B>1)$.
One may therefore use the transfer matrix representation of $\ppt$ to
sum over all spins $n_j'$ except for $j=i$, and over all spins $n_j$ with $j<i-B+1$ or $j>i+B-1$.
The result (for a periodic chain of length $L$) is
\begin{equation}
\fl
F_B = \frac{1}{\mathrm{tr}(M^L)}
       \sum_{n_i'} \sum_{n_{i-B+1},\dots,n_{i+B-1}}  O(\CC,\CC') 
       \left[ \prod_{j=i-B+1}^{i-1} M_{{\cal B}_j,{\cal B}_{j+1}} \right]
       (M^{L-B+1})_{{\cal B}_{i},{\cal B}_{i-B+1}}
                    \label{equ:Fvar-block}
\end{equation}
For long chains, the matrix element $(M^{L-B+1})_{{\cal B},{\cal B}'}$ can be replaced by $\lambda_\mathrm{max}^{L-B+1} x({\cal B}) y({\cal B}')$
where $\lambda_\mathrm{max}$ is the largest eigenvalue of $M$ and $x$ and $y$ the corresponding 
right and left eigenvectors.  The resulting expression
may therefore be evaluated by constructing and diagonalising $M$.  After minimising $F$ over the variational parameters $z_{\cal B}$, one may
then evaluate any one-time observable in the $\nu$-ensemble, via the transfer matrix $M$.

\subsection{Exponential decay of domain distribution in the block model}
\label{app:pd-exp}



Here, we outline a derivation that shows that if the effective potential of a system is a block model of range $B$
then distributions of domain sizes decay exponentially for ranges $r>B$.  For systems described by a block model,
the probability of observing a block in state $\cal B$ is 
\begin{equation}
P({\cal B}) = \frac{ \mathrm{tr}( M^N e^{\cal B})  }{ \mathrm{tr}( M^N ) }
\end{equation}
where $M$ is the transfer matrix of the previous section, and $e^{\cal B}$ is a diagonal matrix with $e_{{\cal B},{\cal B}}=1$ and
all other entries being zero.  Assuming that the matrix $M$ has a gap between its largest and second largest
eigenvalues then for very large $N$, $M^N \approx \bm{v} \lambda^N \bm{u}^T$ where $\lambda$ is the largest eigenvalue
of $M$ and $\bm{u},\bm{v}$ the associated left and right eigenvectors, normalised such that $\bm{u}\cdot\bm{v}=1$.   
Hence, as $N\to\infty$, one has $P({\cal B}) \to v_{\cal B} u_{\cal B}$.

Further, the probability that blocks $i$ and $i+1$ are in states ${\cal B}, {\cal B}'$ is
\begin{equation}
P({\cal B},{\cal B}') = \frac{ \mathrm{tr}( M^{N-1} e^{\cal B} M e^{\mathcal{B}'})  }{ \mathrm{tr}( M^N ) }
\end{equation}
Again, for large $N$ the trace is dominated by the largest eigenvalue of $M$, so that
$P({\cal B},{\cal B}') \to u_{\cal B} M_{{\cal B},{\cal B}'} v_{{\cal B}'} / \lambda$.  
Finally, the analogous property for three successive blocks (for $N\to\infty$) is easily shown to be
$P({\cal B},{\cal B}',{\cal B}'') \to u_{\cal B} M_{{\cal B},{\cal B}'} M_{{\cal B}',{\cal B}''} v_{{\cal B}''} / \lambda^2$
from which we can read off that
\begin{equation}
P({\cal B},{\cal B}',{\cal B}'') = \frac{ P({\cal B},{\cal B}') P({\cal B}',{\cal B}'') }{ P({\cal B}') }.
\label{equ:BBB}
\end{equation}

Recalling that specifying three successive blocks determines the configuration of $B+2$ successive spins, consider
the probability of finding these spins in the configuration $(n_i,0^B,n_{i+B+1})$, where $0^B$ stands for $B$ successive
down spins.  From (\ref{equ:BBB}), this is seen to be
\begin{equation}
P(n_i,0^B,n_{i+B+1}) = \frac{ P(n_i,0^B) \, P(0^B,n_{i+B+1}) }{ P(0^B) },
\end{equation}
and generalising to more than three blocks leads to the more general formula
\begin{equation}
P(n_i,0^{B+x},n_{i+B+x+1}) = \frac{ P(n_i,0^{B}) \, P(0^{B+1})^{x} \, P(0^{B},n_{i+B+x+1}) }{ P(0^{B})^{x+1} }.
\end{equation}
Finally, identifying the domain size distribution
$p(d) = P(1,0^{d-1},1)/P(1)$, one sees that $p(d)$ decays exponentially for $d>B$, proportional
to $[P(0^{B+1})/P(0^B)]^{d-B}$.  A similar result is familiar for the domain structure in one-dimensional Ising systems:
in that case $B=2$, and we note that our definition of $p(d)$ 
is then directly related to the distribution of sizes of spin-down domains.

\subsection{Variational free energy for $p_d$-model}
\label{app:pd}

The calculation of the variational free energy for the $p_d$-model also relies on properties of the matrix
element $O_i(\CC,\CC')$ given in (\ref{equ:Oi}).  Since $O_i(\CC',\CC)\neq 0$ only if $n_{i-1}=1$, one should
consider only configurations $\CC$ where a `domain' starts at site $i-1$.  This occurs with probability 
$\langle n_{i-1}\rangle^\mathrm{var}=\frac{1}{\sum d p_d}$
where we use the notation $\langle\cdot\rangle^\mathrm{var}$ for averages with respect to the trial distribution $\ppt$.  Recalling
in addition that $O_i(\CC',\CC)=0$ unless either $\CC=\CC'$ or $\CC'$ and $\CC$ differ only at spin $i$, one finds that $O_i(\CC',\CC)$
depends on the size $d$ of the domain starting at site $i-1$; if $d>1$ then it depends only on $d$ while if $d=1$ then $O_i(\CC',\CC)$
also depends on the size of domain $d'$ that starts at $i$.  Identifying the relevant cases leads directly from (\ref{equ:Fvar-H})
to (\ref{equ:Fd}).

\subsection{Variational free energy for the $p_{fe}$-model}
\label{app:pfe}

Within the $p_{fe}$ model, the variational free energy is calculated similarly to the $p_d$ model.
The probability that spin $i-1$ occupies a particular position within a domain of parameters $(f,e)$ is 
$p_{fe}/\sum_{f'e'}(f'+e')p_{f'e'}$.  The matrix element $O_i(\CC,\CC')\neq0$ only if  spin $i-1$ is one 
of the $f$ up spins in this domain.  The derivation of the variational free energy then follows that
for the $p_d$ model, except that it requires an additional explicit summation over the $f$ possible positions
of spin $i-1$ within the domain.
The matrix element $O_i(\CC,\CC')$ depends only on the domain containing site $i-1$, except
in the case that this domain has $e=1$, in which case it depends additionally on the next domain
to the right.  Enumerating the specific cases, one arrives at (\ref{equ:Ffe}).

\end{appendix}


\begin{thebibliography}{99}
\bibitem{pi-foot}
To make this approximate scaling relation precise, one may define 
$\pi_K(k) = \lim_{N\to\infty} \lim_{\tobs\to\infty} (-N\tobs)^{-1} \ln P_0(N\tobs k)$: the order and existence of the limits is
discussed, for example, in~\cite{kcm-transition}.
\bibitem{psi-foot}
More
precisely, $(-N\tobs)^{-1}\ln\langle \ee^{-sK} \rangle_0\to\psi_K(s)$ as $N,\tobs\to\infty$:
the relation is analogous to that between $P_0(K)$ and $\pi_K(k)$.
\end{thebibliography}
\end{document}